\documentclass[12pt]{article}
\pdfoutput=1

\usepackage{jheppub}
\usepackage{amssymb,amsfonts,amsmath,amsthm, enumerate}
\usepackage{mathrsfs}
\usepackage{bbold}
\usepackage{tikz}
\usetikzlibrary{shapes.geometric, arrows}
\usepackage{comment}
\usepackage{tikz}
\usepackage{hyperref}
\usepackage{cancel}
\usepackage{soul}

\makeatletter
\newcommand{\Vast}{\bBigg@{4.75}}
\makeatother

\newcommand{\be}{\begin{equation}}
\newcommand{\ee}{\end{equation}}
\newcommand{\bea}{\begin{eqnarray}}
\newcommand{\eea}{\end{eqnarray}}

\newcommand{\scA}{\mathscr{A}}
\newcommand{\scB}{\mathscr{B}}
\newcommand{\scC}{\mathscr{C}}
\newcommand{\scD}{\mathscr{D}}

\newcommand{\CA}{\mathcal{A}}

\newcommand{\CD}{\mathcal{D}}

\newcommand{\CK}{\mathcal{K}}
\newcommand{\CL}{\mathcal{L}}

\newcommand{\CM}{\mathcal{M}}
\newcommand{\CO}{\mathcal{O}}

\newcommand{\CT}{\mathcal{T}}

\newcommand{\CZ}{\mathcal{Z}}

\newcommand{\bq}{{\bf q}}

\newcommand{\lr}{\left (}
\newcommand{\rr}{\right )}
\newcommand{\ls}{\left [}
\newcommand{\rs}{\right ]}
\newcommand{\lc}{\left \{}
\newcommand{\rc}{\right \}}

\newcommand\qt\tau

\newcommand{\p}{\partial}

\renewcommand{\tilde}[1]{\widetilde{#1}}
\newcommand{\tr}{\text{tr}}

\makeatletter
\renewcommand{\@seccntformat}[1]{\csname the#1\endcsname.\,\,}
\makeatother
\allowdisplaybreaks

\let \savenumberline \numberline
\def \numberline#1{\savenumberline{#1.}}

\makeatletter
\def\@fpheader{\relax}
\makeatother

\def\bea{\begin{eqnarray}}
\def\eea{\end{eqnarray}}

\usetikzlibrary{shapes,arrows,chains}
\usetikzlibrary{decorations.markings}
\usetikzlibrary{decorations.pathmorphing}
\usetikzlibrary{shapes.multipart}
\tikzset{snake it/.style={decorate, decoration=snake}}

\newcommand{\zs}{S}
\newcommand{\zm}{M}

\newcommand{\zp}{P}
\newcommand{\zq}{Q}
\newcommand{\zw}{W}
\newcommand{\zv}{V}

\newcommand{\zcw}{\mathcal{W}}
\newcommand{\SG}{E}
\newcommand{\CLG}{G}

\title{\ \vspace{1.6cm} \\
New Heat Kernel Method in Lifshitz Theories}
\author{Kevin T. Grosvenor${}^a$, Charles Melby-Thompson${}^b$, Ziqi Yan${}^{c}$}
\emailAdd{kevinqg1@gmail.com}
\emailAdd{charlesmelby@gmail.com}
\emailAdd{ziqi.yan@su.se}
\affiliation{
${}^a$ Max-Planck-Institut f\"ur Physik komplexer Systeme and \\ W{\"u}rzburg-Dresden Cluster of Excellence ct.qmat \\ N\"othnitzer Str. 38, 01187 Dresden, Germany \medskip\\
${}^b$ 
Institut f\"{u}r Theoretische Physik und Astrophysik,
Julius-Maximilians-Universit\"{a}t W\"{u}rzburg, Am Hubland, 97074 W\"{u}rzburg, Germany \medskip\\
${}^c$ Nordita, KTH Royal Institute of Technology and Stockholm University\\
Roslagstullsbacken 23, SE-106 91 Stockholm, Sweden}
\abstract{We develop a new heat kernel method that is suited for a systematic study of the renormalization group flow in Ho\v{r}ava gravity (and in Lifshitz field theories in general). This method maintains covariance at all stages of the calculation, which is achieved by introducing a generalized Fourier transform covariant with respect to the nonrelativistic background spacetime. As a first test, we apply this method to compute the anisotropic Weyl anomaly for a $(2+1)$-dimensional scalar field theory around a $z=2$ Lifshitz point and corroborate the previously found result. We then proceed to general scalar operators and evaluate their one-loop effective action. The covariant heat kernel method that we develop also directly applies to operators with spin structures in arbitrary dimensions.}% \\

\begin{document}%%%%%%%%%%%%%%%%%%%%%%%%%%%%%%%%%%%%%%%%%%%%%%%%%%%%%%%%%%%%%%%%%%%%%%%%%%%%%%%
%%%%%%%%%%%%%%%%%%%%%%%%%%%%%%%%%%%%%%%%%%%%%%%%%%%%%%%%%%%%%%%%%%%%%%%%%%%%%%%

\maketitle
\vfill\eject

\section{Introduction}

Ho\v{r}ava gravity has attracted much attention over the years since its original proposal in 2009 \cite{Horava:2009uw}. Allowing that Lorentz invariance be absent at high energies, higher spatial derivative terms are introduced in Ho\v{r}ava gravity in such a way that the theory exhibits an anisotropic scaling of space and time. This construction results in a class of unitary and power-counting renormalizable quantum theories of gravity, which are fundamentally nonrelativistic. It is then hoped that the theory is driven to a relativistic fixed point at low energies by classical renormalization group (RG) flow. Moreover, due to the existence of a preferred time direction in Ho\v{r}ava gravity, it admits a rigid notion of causality and its Hamiltonian and diffeomorphism constraints form a closed algebra, at least for some versions of the theory.\,\footnote{This has been demonstrated for the projectable case \cite{Horava:2008ih}, where an extra degree of freedom is present in addition to the ones in General Relativity. For the nonprojectable case, the situation is more intricate (see, e.g., \cite{Bellorin:2010te}).} This is in contrast to the situation in General Relativity, where these constraints fail to form a Lie algebra, leading to intrinsic conceptual and technical difficulties in defining a relativistic quantum gravity nonperturbatively.  

Despite the desirable features of Ho\v{r}ava gravity, it nevertheless still faces the challenge of explaining the vast catalogue of experimental data that highly constrains Lorentz violation \cite{Kostelecky:2008ts}. For example, it would require a great deal of fine-tuning to match the speeds of light for different species of low energy probes \cite{Horava:2011gd}. Different mechanisms for the emergence of low-energy Lorentz invariance have been discussed in the literature (see, e.g., \cite{Coates:2018vub} and references therein). One mechanism is to impose supersymmetry at high energies \cite{GrootNibbelink:2004za}. Another promising mechanism relies on the existence of a strongly-coupled fixed point, in which case it is possible for the strong dynamics to drive the theory sufficiently quickly towards the Lorentzian fixed point \cite{Anber:2011xf, Bednik:2013nxa}. 

Regardless of whether or not Ho\v{r}ava gravity is phenomenologically viable for describing our universe in $3+1$ dimensions, it still has a plethora of important applications in the context of the AdS/CFT correspondence for nonrelativistic systems \cite{Griffin:2011xs, Baggio:2011ha, Griffin:2012qx, Christensen:2013lma, Christensen:2013rfa}, the Causal Dynamical Triangulation approach to quantum gravity \cite{Ambjorn:2013tki, Horava:2009if}, the formulation of membranes at quantum criticality \cite{Horava:2008ih}, the geometric theory of Ricci flow on Riemannian manifolds \cite{Frenkel:2020dic}, and so on. For both phenomenological and formal interests, it is important to understand the quantum structure of Ho\v{r}ava gravity.

In recent years, there has been some intriguing progress made in mapping out the RG flow of the so-called ``projectable'' Ho\v{r}ava gravity, in which the lapse function is independent of space \cite{Barvinsky:2015kil, Barvinsky:2017kob,
Griffin:2017wvh,
Barvinsky:2019rwn, Benedetti:2013pya}. First, the perturbative renormalizability of projectable Ho\v{r}ava gravity has been proven in \cite{Barvinsky:2015kil}.
Then, it was shown in \cite{Barvinsky:2017kob} that there is an asymptotically free ultraviolet (UV) fixed point in $2+1$ dimensions.\footnote{Unlike General Relativity, which is topological in $2+1$ dimensions, Ho\v{r}ava gravity in $2+1$ dimensions has one propagating degree of freedom.} In addition, the anomalous dimension of the cosmological constant in the same theory was evaluated in \cite{Griffin:2017wvh}. Finally, partial results of the RG flow in $3+1$ dimensions were later obtained in \cite{Barvinsky:2019rwn}. 
Nevertheless, knowledge of the RG flow in Ho\v{r}ava gravity without the projectability condition remains rather limited. To reveal this more general quantum structure, which is essential for many phenomenological and formal applications of Ho\v{r}ava gravity, it is urgent to look for more efficient methods. 

One powerful technique for evaluating the one-loop effective action for quantum field theories (QFTs) on a curved background spacetime is the heat kernel method. Different heat kernel methods have been applied in the past to evaluate quantum corrections in QFTs around a Lifshitz fixed point \cite{Nesterov:2010yi, Baggio:2011ha, Mamiya:2013wqa, DOdorico:2015pil, Barvinsky:2017mal}. These heat kernel methods rely heavily on the Zassenhaus formula, which is the inverse companion of the Baker-Campbell-Hausdorff
formula, and in general involves a rather tedious procedure of evaluating nested commutators of covariant time and space derivatives. In some special cases, various simplifications are available. However, when the projectability condition is lacking, the calculation of the full effective action can be inefficient and even computationally impractical.\footnote{For example, \cite{DOdorico:2015pil} utilizes a clever inverse Laplace transform method to express heat kernel coefficients of anisotropic operators, whose spatial derivative term is just a power of the isotropic one, in terms of the heat kernel coefficients of the isotropic operator. However, as a consequence, in this method, one has to perform the Zassenhaus expansion twice: once before the inverse Laplace transform and again afterwards, essentially to undo the expansion in the first place. Except in the case when temporal and spatial covariant derivatives can be chosen to commute, for example if one is interested only in spatial curvature terms in the effective action, this method is technically challenging to implement. Furthermore, it does not apply to anisotropic operators which are not simply powers of isotropic ones, such as the case we study in \S\ref{sec:examples}.}  

We therefore devote this paper to the formulation of a new heat kernel method, which we find better suited to non-projectable cases. We aim to use this new method to systematically study the RG flow in Ho\v{r}ava gravity. In general, this method can be applied to evaluate the one-loop effective action in Lifshitz QFTs on a nonrelativistic background geometry. 

The method we develop in this paper generalizes the algorithm for relativistic theories introduced by Gusynin in \cite{Gusynin:1989ky, gusynin1990seeley}
(also see \cite{gusynin1991asymptotics, Gusynin:1991mk, Gusynin:1990ek, Gorbar:1996xq, Gusynin:1997qs}) to nonrelativistic models. In Gusynin's approach, general covariance is maintained at all stages of the calculation by introducing a generalized Fourier transform that is covariant with respect to the background geometry. This type of covariant Fourier transforms has been studied extensively by mathematicians in the symbolic calculus for pseudodifferential operators \cite{widom1978families, widom1980complete}, which replaces the phase factor in the usual Fourier transform with a phase function that resembles the world function in the Schwinger-DeWitt technique (see \S\ref{sec:cft}). The world function is introduced in the original Schwinger-DeWitt technique to construct an ansatz for the asymptotic expansion of the heat kernel. However, this ansatz is rather special to the minimal second-order differential operator, which complicates its application to more general operators. Nevertheless, Gusynin's approach avoids this intrinsic drawback and has the advantage of being algorithmic for computer programs. 
In this paper we show that there exists a natural generalization of the covariant Fourier transform used in \cite{Gusynin:1989ky} to background geometries equipped with a foliation structure. Using this, we develop a general algorithm for calculating one-loop quantum corrections in Lifshitz theories.

This paper is organized as follows. In \S\ref{sec:chk}, we give a detailed review of the heat kernel method of Gusynin \cite{Gusynin:1989ky, gusynin1990seeley}, which is designed for relativistic QFTs. In \S\ref{sec:chkfs}, we generalize Gusynin's method to Lifshitz field theories. In \S\ref{sec:examples}, we apply the new heat kernel method developed in \S\ref{sec:chkfs} to calculate the one-loop effective action for the most general $z=2$ Lifshitz scalars on a curved background geometry in $2+1$ dimensions, as a first step towards the evaluation of RG flows in nonprojectable Ho\v{r}ava gravity in $2+1$ dimensions. The full result of this calculation is recorded in Appendix \ref{app:a4W}. We conclude our paper in \S\ref{sec:concl}. In Appendix \ref{app:E2}, we show a procedural example that details the evaluation of heat kernel coefficients. In Appendix \ref{app:dis}, we clarify some of the different sign conventions used in the literature to clear up any potential confusion.

\section{Covariant Heat Kernel Method on Riemannian Manifolds} \label{sec:chk}

We start with a review of Gusynin's covariant heat kernel method for quantum field theories on a Riemannian manifold, following closely \cite{Gusynin:1989ky, gusynin1990seeley}.

\subsection{Effective Action and the Heat Kernel Representation}

Consider a QFT on a $d$-dimensional Riemannian manifold $\CM$ equipped with a metric $g_{\mu\nu}$ for a field configuration $\Phi^A (x)$, where $x^\mu$, $\mu = 0, \cdots, d-1$ denotes a set of coordinates on $\CM$.\footnote{More precisely, coordinates on a patch of $\CM$ as part of an atlas on $\CM$.} We take the time to be imaginary.
The field $\Phi^A$ can have a general tensor structure and $A$ denotes collectively its indices. Splitting $\Phi^A$ into its background value $\Phi^A_0$ and the fluctuation $\xi^A$ around this background value, we have the following expansion:
\be
    \Phi^A = \Phi^A_0 + \sqrt{\hbar} \, \xi^A + O (\hbar)\,.
\ee
We assume that the QFT is described by an action principle $S [\Phi]$, which expands as
\begin{align}
    S [\Phi] = S [\Phi_0] & + \sqrt{\hbar} \int \! d^d x \sqrt{g} \, S^{(1)}_A (x) \, \xi^A (x) \notag \\[2pt]
    & + \tfrac{1}{2} \, \hbar \int \! d^d x \sqrt{g} \, \xi^A (x) \, \CO_{AB} \, \xi^B (x) + \CO (\hbar^{3/2})\,,
\end{align}
where $g = \det g_{\mu\nu}$ and
\be
    \sqrt{g} \, S^{(1)}_A (x) = \frac{\delta S[\Phi]}{\delta \Phi^A (x)}\,,
        \qquad
    \sqrt{g} \, \CO_{AB} \, \delta^{(d)}(x-x') = \frac{\delta^2 S[\Phi]}{\delta \Phi^A (x) \, \delta \Phi^B (x')}\,.
\ee
To be specific, we assume that $\Phi^A$ is bosonic. However, the derivation works similarly for fermionic fields.
For concreteness, we set $\CO_{AB}$ to be
\be \label{eq:COAB}
    \CO_{AB} = \sum_{k=1}^{\Delta} \alpha_{AB}^{\mu_1 \cdots \mu_k} (x) \, \nabla_{\mu_1} \cdots \nabla_{\mu_k} + \alpha_{AB} (x)\,,
\ee
where $\Delta$ is a positive even integer and is the order of the differential operator and $\nabla_{\!\mu}$ is the covariant derivative compatible with the metric $g_{\mu\nu}$ (i.e., $\nabla_{\!\mu} \, g_{\nu\rho} = 0$).

The associated (Euclidean) path integral with a source $J_A (x)$ is
\begin{align}
    \mathcal{Z} & = \int d \xi \, \exp \lc -\hbar^{-1} \Bigl[ S [\Phi] - \int \! d^d x \sqrt{g} \, J_A (x) \, \xi^A (x) \Bigr] \rc\,.
\end{align}
Choosing the background value $\Phi_0^A$ such that $S_A^{(1)} = J_A$\,, we obtain that, in the semi-classical limit $\hbar \rightarrow 0$\,, the path integral is approximated by
\be
    \CZ \propto e^{-\hbar^{-1} S [\Phi_0]} \, \bigl( \det \CO_{AB} \bigr)^{-1/2}. 
\ee
We will drop the internal indices $A$ and $B$ in $\CO_{AB}$ in the following.
We therefore read off the quantum effective action,
\be \label{eq:effaction}
    \Gamma [\Phi] = S[\Phi] + \hbar \, \Gamma_1 + O \bigl(\hbar^2\bigr)\,.
\ee
We take the standard heat kernel representation \cite{Vassilevich:2003xt} of the effective action 
\begin{align} \label{eq:hkrep}
    \Gamma_1 & = \frac{1}{2} \, \tr \log \bigl( \CO / \mu^2 \bigr) \notag \\[2pt]
        & = - \frac{1}{2} \frac{d}{ds} \Big|_{s=0} \frac{\mu^{2s}}{\Gamma(s)} \int d^d x \sqrt{g} \int_0^{\infty} d\tau \, \tau^{s-1} \, \CK_\CO (x, x \, | \, \tau)\,.
\end{align}
Note that \eqref{eq:hkrep} is defined after analytically continuing the domain of $s$ to include $s=0$\,. 
We have introduced the heat kernel for the operator $\CO$ in \eqref{eq:hkrep},
\be \label{eq:KO}
    \CK_\CO (x, x_0\,|\,\tau) = \langle x | e^{-\tau \, \CO} |x_0 \rangle\,,
\ee
which satisfies the heat kernel equation $\bigl( \p_\tau + \CO \bigr) \CK_\CO = 0$\,.

Note that the symbol $\CK_{\CO}$ without arguments and the name ``heat kernel'' will be used interchangeably to refer to the operator $e^{- \tau \, \CO}$ in general or to the specific coordinate space representation of this operator, $\CK_{\CO} ( x, x_0 \, | \, \tau )$\,.

\subsection{The Heat Kernel Coefficients} \label{sec:SGc}

The coefficients in the asymptotic expansion of the heat kernel $\CK_\CO$ in the high energy regime $\tau \rightarrow 0^+$ encodes important physical data such as the renormalization group flow in the associated QFT.
There exist different methods in the literature for extracting these heat kernel coefficients.\,\footnote{The asymptotic expansions of the heat kernel were first introduced by DeWitt, which resembles Hadamard's expansion for the retarded Green's function. It was also noted in \cite{Gibbons:1978dw} that the Hadamard-DeWitt expansion is closely related to an expansion given by Minakshisundaram in the study of elliptic operators. Therefore, Gibbons proposed the name Hamidew (Hadamard-Minakshisundaram-DeWitt) for the heat kernel coefficients. The same type of heat kernel expansion was also developed by Seeley and Gilkey in the study of pseudodifferential operators, which we will primarily follow in this paper. To avoid possible confusion regarding the terminology, we will stick to the neutral term ``heat kernel coefficient" in this paper.} One most widely used method was developed by Schwinger and DeWitt, where an ansatz called the Schwinger-DeWitt expansion is taken for the heat kernel~\cite{schwinger1951gauge, dewitt1965dynamical}. The DeWitt expansion ensures the covariance of the calculation of the heat kernel coefficients. The original method of Schwinger and DeWitt is valid for second-order minimal operators, \emph{i.e.} $\CO = - \nabla^\mu \nabla_{\!\mu} + V (x)$\,. This method was then generalized to higher-order nonminimal operators, for example in~\cite{barvinsky1985generalized}. However, this generalization was achieved at the cost of involving nonlocal operators in the intermediate steps of the calculations.\footnote{While the Schwinger-DeWitt method keeps covariance at all stages and invokes recursive methods for solving the heat kernel coefficients, sometimes it is also useful to evaluate the heat kernel in a non-covariant but more direct way, without relying on the recursive methods. This is usually done by using the ordinary plane waves as in Fujikawa's method of calculating the chiral anomaly. See, for example, \cite{PhysRevD.31.3291, PhysRevD.39.1567}.}  

On the other hand, the Seeley-Gilkey method \cite{Seeley:1967ea, Gilkey:1975iq} for calculating the heat kernel coefficients uses techniques developed in the theory of pseudodifferential operators, which avoids introducing nonlocal operators. A pseudodifferential operator $\CO$ is defined by its symbol $\sigma_\CO (x\,, k)$ via
\be \label{eq:Osymbol}
    \langle x | \CO | x_0 \rangle = \int \frac{d^d k}{(2\pi)^d \sqrt{g(x_0)}} \, e^{i k \cdot ( x-x_0)} \, \sigma_\CO (x\,, k)\,.
\ee
We defined $k \cdot x = k_\mu x^\mu$\,.
The heat kernel \eqref{eq:KO} is likewise defined by its symbol $\sigma_{{\CK}_{\CO}}$. We write the heat kernel in terms of its resolvent as
\be
    e^{- \tau \, \CO} = i \int_C \frac{d\lambda}{2 \pi} \, e^{- \tau \, \lambda} \bigl( \CO - \lambda \bigr)^{-1}\,,
\ee
where $C$ is a contour that bounds the spectrum of the operator $\CO$ in the complex plane and is traversed in the counter-clockwise direction. Then, the symbol $\sigma_{{\CK}_{\CO}}$ can be exchanged with the symbol $\sigma_{( \CO - \lambda )^{-1}}$ of the resolvent $( \CO - \lambda )^{-1}$:
\begin{equation} \label{eq:KG}
    \CK_{\CO} (x\,, x_0\,|\,\tau) = i \int_{C} \frac{d \lambda}{2 \pi} \, e^{- \tau \lambda} \, G(x\,, x_0\,|\,\lambda)\,,
\end{equation}
where
\be \label{eq:propagator}
    G(x\,,x_0\,|\,\lambda) \equiv \bigl\langle x \bigl| \bigl( \CO - \lambda \bigr)^{-1} \bigr| x_0 \bigr\rangle = \int \frac{d^d k}{( 2 \pi )^d \sqrt{g(x_0)}}  \, e^{i k \cdot (x-x_0)} \, \sigma (x\,, k \, | \, \lambda)\,.
\ee
Here, we have written $\sigma_{( \CO - \lambda )^{-1}} (x,k)$ as $\sigma ( x,k \, | \, \lambda )$ for ease of notation because this is the only symbol that we will actually care about. Further note that
\be \label{eq:identity}
    \bigl[ \CO (x\,, \nabla) - \lambda \bigr] \, G (x, x_0\,|\,\lambda) = \frac{1}{\sqrt{g(x_0)}} \, \delta^{(d)} (x-x_0)\,.
\ee
Using \eqref{eq:propagator} and \eqref{eq:identity}, we find
\be \label{eq:conjuQ}
    e^{-i k \cdot (x-x_0)} \ls \CO(x\,, \nabla) - \lambda \rs e^{i k \cdot (x-x_0)} \, \sigma (x\,, k \, | \, \lambda) = 1\,. 
\ee

It is well known that the part of the heat kernel diagonal in $x$ has the following asymptotic expansion around $\tau \rightarrow 0^+$~\cite{Seeley:1967ea, Gilkey:1975iq}:
\begin{equation} \label{eq:Seeley}
    \CK_{\CO} (x_0\,,x_0\,|\,\tau) = \sum_{m=0}^{\infty} \SG^{(m)} (x_0) \, \tau^{\frac{m-d}{\Delta}} \,.
\end{equation}
The coefficients $E^{(n)} (x_0)$ are the heat kernel coefficients. Recall that $\Delta$ denotes the order of the operator $\CO$ and is taken to be a positive even integer. To compute the heat kernel coefficients, we expand the symbol $\sigma (x_0\,, k \, | \, \lambda)$ as 
\be \label{eq:expandsigma}
    \sigma = \sum_{m=0}^\infty \sigma^{(m)},
\ee
where $\sigma^{(m)}$ is a homogeneous function of $\lambda$ and $k_\mu$\,, i.e.,
\be \label{eq:homo}
    \sigma^{(m)} (x_0\,, b \, k\, |\,b^{\Delta} \lambda) = b^{-m-\Delta} \, \sigma^{(m)} (x_0\,, k \, | \, \lambda)\,.
\ee
Plugging \eqref{eq:expandsigma} into \eqref{eq:conjuQ} and then taking the rescalings $\lambda \rightarrow b^{\Delta} \, \lambda$ and $k_\mu \rightarrow b \, k_\mu$\,, the coefficients $\sigma^{(m)}$ can be determined recursively by matching terms of different orders in $b$ in the resulting equation. 

Finally, we take the coincidence limit $x_0 \rightarrow x$ in \eqref{eq:KG}, followed by the rescalings 
\be
    \lambda \rightarrow \tau^{-1} \lambda\,,
        \qquad
    k_\mu \rightarrow \tau^{- \frac{1}{\Delta}} k_\mu\,.
\ee
Then, plugging in the expansions \eqref{eq:Seeley} and \eqref{eq:expandsigma} into \eqref{eq:KG} and matching the powers in $\tau$ on both sides of the resulting equation, the heat kernel coefficients are computed to be
\be \label{eq:am}
    \SG^{(m)} (x_0) = \int \frac{d^d k}{(2\pi)^d \sqrt{g(x_0)}} \int_C \frac{i \, d\lambda}{2\pi} \, e^{-\lambda} \, \sigma^{(m)} (x_0\,, k \, | \, \lambda)\,.
\ee
When $m$ is odd, $\sigma^{(m)}$ is odd in $k_\mu$\,, and the corresponding $E^{(m)}$ vanishes. Therefore, $E^{(m)}$ is only nonzero when $m$ is even, and
\begin{equation}
    \CK_{\CO} (x_0\,,x_0\,|\,\tau) = \sum_{r=0}^{\infty} \SG^{(2r)} (x_0) \, \tau^{\frac{2r-d}{\Delta}} \,.
\end{equation}

Note that the phase factor $k \cdot (x-x_0)$ in \eqref{eq:propagator} is not invariant with respect to general coordinate transformations. Therefore, while the heat kernel coefficients are covariant with respect to general coordinate transformations in $x$, the definition of the symbol $\sigma$ in \eqref{eq:propagator} is not. We will now describe Widom's solution \cite{widom1978families, widom1980complete}, which leads to a covariant definition of the symbol $\sigma$, and its application by Gusynin in \cite{Gusynin:1989ky} to the computation of heat kernel coefficients. 

\subsection{Covariant Fourier Transform} \label{sec:cft}

As pointed out by Gusynin in \cite{Gusynin:1989ky}, there is a covariant generalization of the Seeley-Gilkey method 
by applying the covariant Fourier transform as in \cite{widom1978families, widom1980complete}. The idea is to generalize the phase factor $k \cdot (x-x_0)$ to a real-valued phase function
\begin{equation} \label{eq:phf}
    \ell : \CM \times T^* (\CM) \rightarrow \mathbb{R}\,,
\end{equation}
such that $\ell(x\,; x_0\,, k)$ is linear in $k \equiv k_\mu \, dx^\mu \in T^*_{x_0} (\CM)$ for any fixed $x_0\,, x \in \CM$\,. In contrast, there is no canonical meaning to linearity in the variable $x \in \CM$\,. Widom proposed in \cite{widom1978families, widom1980complete} the following definition
for linearity of $\ell$ in $x^\mu$ in the vicinity of the point $x_0$\,: for each $k \in T^*_{x_0} (\CM)$\,, the symmetrized $n$-th covariant derivative vanishes at $x_0$ for $n \neq 1$\,, namely, 
\begin{subequations} \label{eq:linearinx}
\begin{align}
    \nabla_{\!\mu} \ell (x\,; x_0\,, k) \big|_{x=x_0} & = k_\mu\,, \\[2pt]
    \nabla_{\!(\mu_1} \cdots \nabla_{\!\mu_n)} \ell (x\,; x_0\,, k) \big|_{x=x_0} & = 0\,,
        \,\,\,\,
    n = 0\,, 2\,, 3\,, \cdots\,.
\end{align}
\end{subequations}
It is understood that $\nabla_\mu$ acts on $x$\,. Here, $\nabla_{(\mu_1} \cdots \nabla_{\mu_n)}$ has all the indices symmetrized. For example, $T_{(\mu} {}_{\nu)} = \tfrac{1}{2} \, \bigl( T_{\mu} {}_{\nu} + T_{\nu} {}_{\mu} \bigr)$ for any tensor $T_{\mu\nu}$\,. 
Using \eqref{eq:linearinx}, higher derivatives of the phase function $\ell (x ; x_0 , k)$ with respect to $x$ can be expressed in terms of fewer derivatives at $x=x_0$ and thus all covariant derivatives of $\ell$ are also determined in the coincidence limit $x \rightarrow x_0$\,. Such an $\ell \in C^\infty (\CM)$ always exists in a neighborhood of $x_0$ due to Borel's theorem \cite{widom1980complete}.

The above definition of the phase function is designed such that a covariant generalization of the Taylor series can be constructed. 
Since $\ell (x\,; x_0\,, k)$ is linear in $k$ for fixed $x_0$ and $x$\,, we have
\be
    \ell (x\,; x_0\,, k) = k_\mu \, \ell^\mu (x\,, x_0)\,,
\ee
where $\ell^\mu (x\,, x_0)$ defines a tangent vector. It is then straightforward to show by using \eqref{eq:linearinx} that, for a given function $f(x) \in C^\infty (\CM)$ \cite{widom1980complete}, 
\be \label{eq:WidomTaylor}
    f(x) = \sum_{n=0}^\infty \frac{1}{n!} \nabla_{\!\mu_1} \cdots \nabla_{\!\mu_n} f (x_0) \, \ell^{\mu_1} (x\,, x_0) \cdots \ell^{\mu_n} (x\,, x_0)\,.
\ee
for $x$ in a neighborhood of $x_0$\,.\footnote{More precisely, \eqref{eq:WidomTaylor} holds for the \emph{analytic germ} of the function $f(x)$\,. We take this for granted throughout this paper.} This generalizes the usual Taylor theorem, replacing $\p_\mu$ with the covariant derivative $\nabla_{\!\mu}$\,.

With the phase function $\ell$ in hand, we are able to replace the expressions in \S\ref{sec:SGc} with their covariant counterparts, following \cite{Gusynin:1989ky}.
First, a pseudodifferential operator $\CO$ is now related to its symbol $\sigma_\CO (k\,, x_0\,; x)$ via
\be
    \langle x | \CO | x_0 \rangle = \int \frac{d^d k}{(2\pi)^d \sqrt{g(x_0)}} \, e^{i \, \ell (k, \, x_0; \, x)} \, \sigma_\CO (x\,; x_0\,, k)\,,
\ee
which replaces \eqref{eq:Osymbol} and defines a covariant Fourier transform. It then follows that the matrix element of the operator $\CO$ resolvent introduced in \eqref{eq:propagator} is replaced with
\be \label{eq:propagatorcov}
    G(x\,,x_0\,|\,\lambda) \equiv \bigl\langle x \bigl| \bigl( \CO - \lambda \bigr)^{-1} \bigr| x_0 \bigr\rangle = \int \frac{d^d k}{( 2 \pi )^d \sqrt{g(x_0)}}  \, e^{i \, \ell (x; \, x_0, \, k)} \, \sigma (x\,; x_0\,, k\,|\,\lambda)\,.
\ee
We defined $\sigma (x\,; x_0\,, k\,|\,\lambda) \equiv \sigma_{( \CO - \lambda )^{-1}} (x\,; x_0\,, k)$\,. The heat kernel $\CK_\CO$ takes the same form as in \eqref{eq:KG}, i.e.,
\begin{equation} 
    \CK_{\CO} (x\,, x_0\,|\,\tau) = i \int_{C} \frac{d \lambda}{2 \pi} \, e^{- \tau \lambda} \, G(x\,, x_0\,|\,\lambda)\,.
\end{equation}

Identical to \eqref{eq:identity}, we have
\be \label{eq:identity2}
    \bigl[ \CO (x\,, \nabla) - \lambda \bigr] \, G (x, x_0\,|\,\lambda) = \frac{1}{\sqrt{g(x_0)}} \, \delta^{(d)} (x-x_0)\,,
\ee
but with $G$ represented as in \eqref{eq:propagatorcov}. For \eqref{eq:identity2} to be satisfied, we require the following analogue of \eqref{eq:conjuQ}:
\be \label{eq:Qll}
    \bigl[ \CO (x\,, \nabla \! + i \nabla \ell) - \lambda \bigr] \, \sigma (x\,; x_0\,, k\,|\,\lambda) = I (x\,, x_0)\,,
\ee
where the biscalar $I(x\,, x_0)$ satisfies
\be \label{eq:Idelta}
    \int \! \frac{d^d k}{(2\pi)^d} \, e^{i \, \ell (x;\, x_0,\, k)} \, I (x\,, x_0) = \delta^{(d)} (x - x_0)\,.
\ee

There are a couple of things to keep in mind. Firstly, $\CO$ may depend on the fields and their derivatives. The derivatives acting on fields are \emph{not} shifted by $\nabla \rightarrow \nabla + i \nabla \ell$ as in \eqref{eq:Qll}. Only the derivatives which act on $G$ are shifted in this way; these are shifted for the simple reason that when a derivative acts on $G$, it acts on the symbol $\sigma$ as well as the phase factor $e^{i \ell}$ and pulls down a factor of $i \nabla \ell$ from $e^{i\ell}$.

A second point to keep in mind is that $\CO$ implicitly carries the bundle space index structure and should be written as $\CO_{AB}$\,. Therefore, $I$ also carries the index structure $I_{AB} (x\,, x_0)$\,. The covariant derivative $\nabla_{\!\mu}$ then also picks up the bundle space indices and reads $\bigl( \nabla_{\!\mu} \bigr){}_A{}^B$. For example, when Yang-Mills theory on a curved background is under consideration, the index $A$ 
contains both a spacetime index $\mu$ (the vector index of the gauge field) and a gauge group index $a$\,. We have 
\be
    \bigl( \nabla_{\rho} \bigr){}_{\mu a}{}^{\nu b} = \delta_\mu^\nu \, \bigl[ \delta_a^b \, \nabla_\rho - i \, ( t^c )_a{}^b \, \CA_\rho^c \bigr]\,,
\ee
where $t^a$ are generators in the gauge group and $\CA_\mu^a$ the gauge field. From \eqref{eq:Idelta} and in analogue of \eqref{eq:linearinx}, as in \cite{Gusynin:1989ky}, we require that $I_{AB}$ satisfy
\begin{subequations} \label{eq:Idef}
\begin{align}
    I_{AB} (x_0\,, x_0) & = \mathbb{1}_{AB}\,, \\[2pt]
    \bigl(\nabla_{\!(\mu_1}\bigr){}_{A_1}{}^{\!B_1} \cdots \bigl(\nabla_{\!\mu_{n})}\bigr){}_{A_n}{}^{\!B_{n}} \, I_{B_{n} C} (x\,, x_0) \, \big|_{x=x_0} \! & = 0\,,
        \,\,\,\,
    n \geq 1\,.
\end{align}
\end{subequations}
Here, $\mathbb{1}_{AB}$ consists of Kronecker symbols with the bundle index $A$\,. The covariant derivatives only act on the first bundle index in $I_{AB}$\,. The equations in \eqref{eq:Idef} fully determine the coincidence limit $x \rightarrow x_0$ of any number of covariant derivatives acting on $I$\,.

Finally, we take the expansion of $\sigma$ as in \eqref{eq:expandsigma},
\be
    \sigma (x\,; x_0\,, k\,|\,\lambda) = \sum_{m=0}^\infty \sigma^{(m)} (x\,; x_0\,, k\,|\,\lambda)\,,
\ee
where the coincidence limit of $\sigma^{(m)}$ satisfies \eqref{eq:homo} and can be determined recursively. 
The heat kernel coefficients are then determined as in \eqref{eq:am}.

\section{Covariant Heat Kernel Method on Foliated Spacetimes} \label{sec:chkfs}

We will formulate a generalization of the covariant Fourier transform as well as the algorithm for calculating heat kernel coefficients in cases when the spacetime carries a foliation structure. Here, we will consider a foliation structure in which the leaves of the foliation have codimension one, but the discussion can be naturally generalized to other cases. We start with introducing a few essential ingredients about foliated spacetimes that will be useful for subsequent discussion.

\subsection{Geometry of Foliated Spacetimes}

\quad $\bullet$ \textbf{Metric Decomposition} 

\vspace{2mm}

Consider the geometry of a $(D+1)$-dimensional spacetime manifold $\CM$\,. 
Let $\CM$ be foliated by leaves $\Sigma$ of codimension one, which means that $\CM$ is equipped with an atlas of coordinate systems $x^\mu = (t\,, x^i)$\,, $i = 1\,, \cdots\,, D$\,. The transition functions are restricted to be foliation preserving, such that
\be
    \tilde{t} = \tilde{t} (t)\,,
        \qquad
    \tilde{x}^i = \tilde{x}^i (t\,, \mathbf{x})\,.
\ee
This foliated spacetime is naturally described by a rank $D$ degenerate symmetric tensor $h_{\mu\nu}$ together with a choice of a vector $n^{\mu}$ such that $h_{\mu\nu} n^{\nu} = 0$\,. Given this data, there exist unique objects $h^{\mu\nu}$ and $n_{\mu}$ that satisfy the relations\,\footnote{Here, we choose an ``all positive'' convention. However, note that another common convention is to choose $n_{\mu} n^{\mu} = -1$\,, in which case $h^{\mu\rho} h_{\rho\nu} - n^{\mu} n_{\nu} = \delta^{\mu}_{\nu}$\,.}
\begin{align}
    n_{\mu} n^{\mu} &= 1\,, &%
    h^{\mu\nu} n_{\nu} &= 0\,, &%
    h^{\mu\rho} h_{\rho\nu} + n^{\mu} n_{\nu} &= \delta^{\mu}_{\nu}\,.
\end{align}
It is important to keep in mind that $n_{\mu}$ is not retrieved from $n^{\mu}$ by ``lowering'' with respect to any metric. For one thing, there is no actual metric, and secondly, if we were to ``lower'' the index on $n^{\mu}$ using $h_{\mu\nu}$\,, the result would be zero. Instead, $h^{\mu\nu}$ and $n_{\mu}$ are simply objects which are \emph{defined} by the above relations. Define a two-tensor field $g_{\mu\nu}$ and its inverse $g^{\mu\nu}$ over $\CM$ as follows:
\be \label{eq:decomg}
    g_{\mu\nu} = n_\mu n_\nu + h_{\mu\nu}\,,
        \qquad
    g^{\mu\nu} = n^\mu n^\nu + h^{\mu\nu}\,.
\ee
Note that $g_{\mu\nu}$ does not always qualify as a metric field.\footnote{One famous example in which $g_{\mu\nu}$, assembled in this way from $n^{\mu}$ and $h_{\mu\nu}$, does not form a bona fide metric field is realized in Newton-Cartan theory \cite{Misner:1974qy}.}

Frequently, it is convenient to refer to these objects without reference to a coordinate system. For this purpose, we introduce the tensor notation 
\begin{subequations}
\begin{align}
    \overline{h} & = h^{\mu\nu} \, \p_\mu \otimes \p_\nu\,, 
            &% 
    \overline{n} & = n^\mu \, \p_\mu\,, \\[2pt]
    \underline{h} & = h_{\mu\nu} \, dx^\mu \otimes dx^\nu\,, 
        &%
    \underline{n} & = n_\mu \, dx^\mu\,,
\end{align}
\end{subequations}
where $\p_\mu \in T_{x}(\CM)$ and $dx^\mu \in T^*_x(\CM)$\,. Moreover, $\p_\mu (dx^\nu) = dx^\nu (\p_\mu) = \delta^\nu_\mu$\,. The overline (underline) indicates that the object is a $(n,0)$-tensor ($(0, n)$-tensor) living in the tensor product of $n$ tangent (cotangent) spaces.
In this notation, the ``givens'' are $\underline{h}$ and $\overline{n}$ with relation $\underline{h} ( .\,, \overline{n} ) = 0$\,. The ``derived quantities'' are $\overline{h}$ and $\underline{n}$\,, satisfying the relations
\begin{align}
    \underline{n} ( \overline{n} ) &= 1\,, &%
    \overline{h} ( .\,, \underline{n} ) &= 0\,, &%
    \overline{h} ( .\,, \underline{h} ) + \overline{n} \otimes \underline{n} & = \mathbb{1}_{1,1}\,,
\end{align}
where in $\overline{h} (.\,, \underline{h})$ only the first cotangent space over which the tensor product in $\underline{h}$ takes is acted on. Moreover, the (1,1)-tensor $\mathbb{1}_{1,1}$ is $\p_\mu \otimes dx^\mu$ when a coordinate system is chosen.
Similarly, the two-tensor $g_{\mu\nu}$ and its inverse $g^{\mu\nu}$ are associated with the quantities $\underline{g}$ and with its inverse $\overline{g}$\,, respectively, satisfying $\overline{g} (.\,, \underline{g}) = \mathbb{1}_{1,1}$\,. Then, we have \eqref{eq:decomg} cast in the coordinate independent form
\be \label{eq:gg}
    \underline{g} = \underline{n} \otimes \underline{n} + \underline{h}\,,
        \qquad
    \overline{g} = \overline{n} \otimes \overline{n} + \overline{h}\,.
\ee
Finally, $\underline{n}$ is assumed to be orthogonal to the hypersurface $\Sigma$\,, which means that there exists an acceleration vector field $\underline{a} \in T^* (\CM)$ with $\underline{a} ( \overline{n} ) = 0$ such that
\begin{equation}
    d\underline{n} + \underline{a} \wedge \underline{n} = 0\,.
\end{equation}

Now, let us pick a convenient set of coordinates adapted to the foliation structure. Choose a holonomic basis $\{ \overline{e}_i \}$ for $T_x (\Sigma) \subset T_x (\CM)$\,, satisfying
\begin{align}
    [ \overline{e}_i , \overline{e}_j ] &= 0\,, &%
    \underline{n} ( \overline{e}_i ) &= 0\,.
\end{align}
Note that the index $i$ is a label that runs from 1 to $D$ and it does \emph{not} denote the coordinate components of each basis element. The dual basis $\{ \underline{\omega}^i \}$ (generally non-holonomic) is defined by the relations
\begin{align}
    \underline{\omega}^i ( \overline{e}_j ) &= \delta^{i}_{j}\,, &%
    \underline{\omega}^i ( \overline{n} ) &= 0\,.
\end{align}
The spatial metric in this basis is
\begin{equation}
    h_{ij} \equiv \underline{h} ( \overline{e}_i , \overline{e}_j )\,.
\end{equation}
Then, $\{ \overline{e}_{\mu} \} = \{ \overline{n} , \overline{e}_i \}$ and $\{ \underline{\omega}^{\mu} \} = \{ \underline{n} , \underline{\omega}^i \}$ serve as bases for $T( \CM )$ and $T^* ( \CM )$\,, respectively. In coordinates $(t, x^i )$ adapted to this basis, we have
\begin{subequations} \label{eq:lpsh}
\begin{align}
    \underline{\omega}^i &= dx^i + N^i dt\, &%
    \underline{n} &= N \, dt\,, &%
    \underline{h} & = h_{ij} \, \underline{\omega}^i \otimes \underline{\omega}^j\,, \\[2pt]
    \overline{e}_i &= \partial_i\, &%
    \overline{n} &= \frac{1}{N} ( \partial_t - N^i \partial_i )\,, &%
    \overline{h} & = h^{ij} \, \overline{e}_i \otimes \overline{e}_j\,,
\end{align}
\end{subequations}
where $h^{ik} h_{kj} = \delta^i_j$\, and where $N$ and $N^i$ are quantities defined by the above equations. In the case when $\underline{g}$ is a bona fide metric field, $N$ and $N^i$ are, respectively, the \emph{lapse function} and \emph{shift vector} of the ADM metric decomposition and it is customary, even in the Newton-Cartan literature, to continue referring to these quantities as the lapse and shift.

\vspace{2mm}

\noindent \quad $\bullet$ \textbf{Covariant Derivatives} 

\vspace{2mm}

We introduce the Levi-Civita connection $D$ with respect to the two-tensor $\underline{g}$\,. We also require that the manifold is torsionless: for any vector fields $\overline{X}\,, \overline{Y} \in T (\CM)$\,, the torsion tensor $D_{\overline{X}} \, \overline{Y} - D_{\overline{Y}} \, \overline{X} - [\overline{X}\,, \overline{Y}]$ is set to zero. Define the Christoffel symbols via $D_{\overline{e}_\mu} \overline{e}_\nu = \overline{e}_\lambda \, \Gamma^\lambda{}_{\mu\nu}$\,. The Christoffel coefficients can be computed by noting the identity
\begin{align}
    2 \, \overline{g} \bigl( D_{\overline{X}} \overline{Y}, \overline{Z} \bigr) & = \overline{X} \bigl( \overline{g}(\overline{Y}, \overline{Z}) \bigr) + Y \bigl( \overline{g}(\overline{X}, \overline{Z}) \bigr) - \overline{Z} \bigl( \overline{g}(\overline{X}, \overline{Y}) \bigr) \notag \\[2pt]
    & \quad + \overline{g} \bigl( [\overline{X}, \overline{Y}]\,, \overline{Z} \bigr) + \overline{g} \bigl( [\overline{Z}, \overline{X}]\,, \overline{Y} \bigr) + \overline{g} \bigl( [\overline{Z}, \overline{Y}]\,, \overline{X} \bigr)\,,
\end{align}
where the vector fields $\overline{X}, \overline{Y}, \overline{Z}$ can run through $\{ \overline{n}\,, \overline{e}_i \}$\,. This yields, in components,
\begin{equation}
\begin{aligned}
    \Gamma^{k}{}_{ij} &\equiv \gamma^{k}{}_{ij}\,, &%
    \Gamma^{i}{}_{jn} &= K^{i}{}_{j}\,, &%
    \Gamma^{i}{}_{nj} &= K^{i}{}_{j} + L^{i}{}_{j}\,, &%
    \Gamma^{i}{}_{nn} &= a^i\,, \\
    \Gamma^{n}{}_{ij} &= - K_{ij}\,, &%
    \Gamma^{n}{}_{jn} &= 0\, &%
    \Gamma^{n}{}_{nj} &= -a_j\,, &%
    \Gamma^{n}{}_{nn} &= 0\,,
\end{aligned}
\end{equation}
where $\gamma^{k}{}_{ij}$ denotes the Levi-Civita connection of $h_{ij}$ on $T ( \Sigma )$, the Latin indices are raised and lowered by $h_{ij}$ and $h^{ij}$, and
\begin{align}
    K_{ij} &= \frac{1}{2N} ( \dot{h}_{ij} - \nabla_{\!i} N_j - \nabla_{\!j} N_i )\,, &%
    L^{i}{}_{j} &= \frac{\partial_{\!j} N^i}{N}\,, &%
    a_i &= - \frac{\partial_{\!i} N}{N}\,,
\end{align}
are the extrinsic curvature of $\Sigma$ in $\CM$, the shift variation, and the acceleration vector, respectively. Here, $\nabla_{\!i}$ is the covariant derivative with respect to $h_{ij}$\,. The spatial Riemann curvature tensor is defined as usual:
\be \label{eq:riemanntensor}
    R^{i}{}_{jk \ell} = \partial_k \Gamma^{i}{}_{\ell j} - \partial_{\ell} \Gamma^{i}{}_{kj} + \Gamma^{i}{}_{km} \Gamma^{m}{}_{\ell j} - \Gamma^{i}{}_{\ell m} \Gamma^{m}{}_{kj}.
\ee
The Ricci tensor is defined as $R_{ij} = R^{k}{}_{ikj}$ and the Ricci scalar is $R = R_{i}{}^{i}$.

We spare a few words on the geometrical meaning of $L^i{}_j$\,. First note that the torsionless condition $0 = \nabla_{\overline{n}} \, \overline{e}_i - \nabla_{\overline{e}_i} \overline{n} - [ \overline{n} , \overline{e}_i ]$ implies $[ \overline{n} , \overline{e}_i ] = \overline{e}_j L^{j}{}_{i} - \overline{n} \, a_i$\,.
Therefore,
\begin{equation}
    \overline{e}_i L^{i}{}_{j} = \Pi ( [ \overline{n} , \overline{e}_j ] )\,,
\end{equation}
where $\Pi$ is the projection operator from $T ( \CM )$ to $T ( \Sigma )$ that distributes over tensor products, with
$\Pi ( \overline{e}_{\mu} ) = \delta_{\mu}^{i} \, \overline{e}_i$\,.
This implies that $L^{i}{}_{j}$ encodes the action of infinitesimal diffeomorphisms by $\overline{n}$ on $T ( \Sigma )$. %

Now, we define covariant derivatives adapted to the foliation. We have already introduced the ``covariant spatial derivative" $\nabla_{\!i}$\,, which is related to $D$ acting on $\CM$ as
\be
    \nabla_{\!i} \CT \equiv \Pi (D_{\overline{e}_i} \CT)\,,
\ee
for an arbitrary tensor field $\CT$.
On a $C^\infty$ function, $\nabla_i$ acts as $\p_i$\,.
We also need a notion of ``covariant time derivative,'' which we denote by $d_n$\,. For a $C^\infty$ function $f$\,, we simply take 
\be
    d_n(f) \equiv \overline{n} (f) = n^{\mu} \partial_{\mu} f\,;
\ee
for a given spatial tensor field $\CT$\,, it is convenient to define its covariant time derivative to be the Lie derivative with respect to $\overline{n}$\,, projected onto $\Sigma$\,, i.e.,
\be \label{eq:dnT}
    d_n \CT \equiv \Pi \bigl( \CL_{\overline{n}} \CT \bigr) = \Pi \bigl( [\overline{n}\,, \CT] \bigr)\,.
\ee

\vspace{2mm}

\noindent \quad $\bullet$ \textbf{Expressions in Components} 

\vspace{2mm}

For practical calculations, we will eventually need to write various derivatives acting on tensor fields in component form. To facilitate such calculations, we note that $d_n$ and $\nabla_{\!i}$ act on the (co)tangent space basis as
\begin{subequations} \label{eq:neo}
\begin{align}
    \nabla_{\!i}\, \overline{e}_j & = \gamma^k{}_{ij} \, \overline{e}_k\,,
        &
    d_n \overline{e}_i & = L^j{}_i \, \overline{e}_j\,, \\[2pt]
    \nabla_{\!i} \, \underline{\omega}^j & = - \gamma^j{}_{ik} \, \underline{\omega}^k\,,
        &
    d_n \underline{\omega}^i & = - L^i{}_j \, \underline{\omega}^j\,.
\end{align}
\end{subequations}
Furthermore, the commutator $[d_n\,, \nabla_i]$ acts on the same bases as
\begin{subequations} \label{eq:neocomm}
\begin{align}
    [d_n\,, \underline{\omega}^j \, \nabla_{\!j}] \, \overline{e}_i \, & = \bigl( - a_j \, d_n \overline{e}_i + M^k{}_{ij} \, \overline{e}_k \bigr) \otimes \underline{\omega}^j\,, \\[2pt]
    [d_n\,, \underline{\omega}^j \, \nabla_{\!j}] \, \underline{\omega}^k \! & = \bigl( - a_j \, d_n \underline{\omega}_i - M^k{}_{ij} \, \underline{\omega}^i \bigr) \otimes \underline{\omega}^j\,,
\end{align}
\end{subequations}
where the (1,2)-tensor $M$ is defined for any vector fields $\overline{X}$ and $\overline{Y}$ as  
\be
    M(\overline{X}\,, \overline{Y}) \equiv [\overline{n}\,, \nabla_{\overline{X}}] \overline{Y} - \nabla_{\!d_n \overline{X}} \overline{Y} + \underline{a} (\overline{X}) \, d_n \overline{Y}\,.
\ee
In components, we find
\begin{equation}
    M_{kij} = ( \nabla_{\!i} - a_i ) K_{jk} + ( \nabla_{\!j} - a_j ) K_{ik} - ( \nabla_{\!k} - a_k ) K_{ij}\,,
\end{equation}
where we have lowered the first index on $M$ using $h_{ij}$ for neatness.

Using \eqref{eq:neo} and \eqref{eq:neocomm}, it follows that, in components,\footnote{For examples of \eqref{eq:dnTij}, we have $d_n h_{ij} = 2K_{ij}$ and $d_n h^{ij} = - 2K^{ij}$\,.}
\begin{subequations} \label{eq:ddT}
\begin{align}
    \nabla_k \CT^{i_1 \cdots i_m}{}_{j_1 \cdots j_n} = \partial_{k} \CT^{i_1 \cdots i_m}{}_{j_1 \cdots j_n} &+ \sum_{r=1}^{m} \Gamma^{i_r}{}_{k\ell} \, \CT^{i_1 \cdots i_{r-1} \ell i_{r+1} \cdots i_m}{}_{j_1 \cdots j_n} \notag \\
    &- \sum_{r=1}^{n} \Gamma^{\ell}{}_{k j_r} \, \CT^{i_1 \cdots i_m}{}_{j_1 \cdots j_{r-1} \ell j_{r+1} \cdots j_n}\,, \\[4pt]
    d_n \CT^{i_1 \cdots i_m}{}_{j_1 \cdots j_n} = n^{\mu} \partial_{\mu} \CT^{i_1 \cdots i_m}{}_{j_1 \cdots j_n} &+ \sum_{r=1}^{m} L^{i_r}{}_{\ell} \, \CT^{i_1 \cdots i_{r-1} \ell i_{r+1} \cdots i_m}{}_{j_1 \cdots j_n} \notag \\
    &- \sum_{r=1}^{n} L^{\ell}{}_{j_r} \, \CT^{i_1 \cdots i_m}{}_{j_1 \cdots j_{r-1} \ell j_{r+1} \cdots j_n}\,, \label{eq:dnTij} \\[4pt]
    [d_n , \nabla_k] \CT^{i_1 \cdots i_m}{}_{j_1 \cdots j_n} = -a_k \, d_n T^{i_1 \cdots i_m}{}_{j_1 \cdots j_n} &+ \sum_{r=1}^{m} M^{i_r}{}_{k \ell} \, T^{i_1 \cdots i_{r-1} \ell i_{r+1} \cdots i_m}{}_{j_1 \cdots j_n} \notag \\
    &- \sum_{r=1}^{n} M^{\ell}{}_{k j_r} \, T^{i_1 \cdots i_m}{}_{j_1 \cdots j_{r-1} \ell j_{r+1} \cdots j_n}\,. \label{eq:dnnabcomm}
\end{align}
\end{subequations}

\subsection{Covariant Fourier Transform of Anisotropic Operators} \label{sec:cftao}

We have learned from \S\ref{sec:chk} that it is key to introduce a covariant Fourier transform if one desires to compute the heat kernel coefficients in a fully covariant way. To construct a covariant Fourier transform that is natural on a foliated spacetime background, we first need to find a phase function that generalizes the phase factor $k_\mu (x^\mu - x^\mu_0)$ and that is adapted to the foliation. The phase function $\ell$ defined in \eqref{eq:phf} does not qualify as an appropriate choice since it treats all coordinates equally, leading to differential symbols that do not naturally decompose with respect to the foliation. A phase function that is designed to already take into account the foliation structure will significantly simplify the computation of heat kernel coefficients.

Recall that $x^\mu = (t\,, x^i)$\,, $i = 1\,, \cdots, D$\,. In the following, we use $x$ to denote $x^\mu$ and $\mathbf{x}$ to denote $x^i$\,. We start with defining a temporal phase function $\chi (x\,; x_0\,, \underline{\nu}) \in \mathbb{R}$ that generalizes the flat limit phase factor $\nu \, (t-t_0)$\,, with $\nu$ the frequency. We defined
\be \label{eq:nubar}
    \underline{\nu} = \nu \, \underline{n} \in T^*_{x_0} (\CM)\,.
\ee
Note that in the coordinates $(t, x^i )$ adapted to the foliation, $\underline{n} = N \, dt$, and so the coefficient of $dt$ in $\underline{\nu}$ is $N \nu$, not just $\nu$. We have to be careful in dealing with this fact in the derivations that follow. We will mention this again at crucial points.

We require that $\chi(x \,; x_0\,, \underline{\nu})$ be linear in $\underline{\nu}$ for fixed $x_0\,, x \in \CM$\,. It is natural to require $\chi$ to be constant on each spatial slice, which implies that all spatial derivatives vanish identically, i.e.,
\be \label{eq:sdchi}
    \nabla_{\!i_1} \cdots \nabla_{\!i_k} \chi (x\,; x_0\,, \underline{\nu}) = 0\,,
        \qquad
    k \geq 1\,.
\ee
It then follows that
\be \label{eq:chi1}
    d_n^\ell \nabla_{\!i_1} \cdots \nabla_{\!i_k} \chi (x\,; x_0\,, \underline{\nu}) = 0\,,
        \qquad
    k \geq 1\,, \ell \geq 0\,.
\ee
Together with \eqref{eq:dnnabcomm}, these relations imply that $\CD \nabla_{\!i} \, \chi = 0$ for any operator $\CD$ built out of $d_n$'s and $\nabla_{\!j}$'s, which constitutes a significant simplification.

Moreover, we require that $\chi$ satisfy
\begin{subequations} \label{eq:tdchi}
\begin{align}
    d_n \chi (x\,; x_0\,, \underline{\nu}) \big|_{x = x_0} \! & = \nu\,, \label{eq:dnc} \\[2pt]
    d_n^k \chi (x\,; x_0\,, \underline{\nu}) \big|_{x = x_0} \! & = 0\,, \,\,\,\, k = 0\,, 2\,, 3\,, \cdots\,,
\end{align}
\end{subequations}
such that $\chi$ reduces to $\nu(x-x_0)$ in the flat limit. The conditions in \eqref{eq:sdchi} and \eqref{eq:tdchi} determine all $d_n$ and $\nabla_{\!i}$ derivatives of $\chi$ in the coincidence limit $x \rightarrow x_0$\,. The existence of such a function $\chi$ in a neighborhood of $x_0$ is ensured by Borel's theorem.

Next, we define a spatial phase function $\psi (x\,; x_0\,, \underline{q}) \in \mathbb{R}$ that generalizes the flat limit phase factor $q_i (x^i - x^i_0)$\,, with $q_i$ the spatial momentum. We define
\be
    \underline{q} = q_i \, \underline{\omega}^i\in T^*_{x_0} (\CM)\,.
\ee
We require that $\psi (x \, ; x_0 \, , \underline{q})$ be linear in $\underline{q}$ for fixed $x_0\,, x \in \CM$\,. On the spatial slice, we can take the same conditions as in \eqref{eq:linearinx} for $\ell$\,, with
\begin{subequations} \label{eq:dpsiall}
\begin{align}
    \nabla_{\!i} \psi (x\,; x_0\,, \underline{q}) \big|_{x=x_0} & = q_i\,, \label{eq:dip} \\[2pt]
    \nabla_{\!(i_1} \cdots \nabla_{\!i_k)} \psi (x\,; x_0\,, \underline{q}) \big|_{x=x_0} & = 0\,,
        \,\,\,\,
    k = 0\,, 2\,, 3\,, \cdots\,. \label{eq:dpc}
\end{align}
\end{subequations}
We further require that these conditions on $\psi$ be trivially covariantly transported from one leaf of the foliation to the next (i.e., by $d_n$), by demanding that
\be
    d_n^\ell \nabla_{(i_1} \cdots \nabla_{i_k)} \psi(x\,; x_0\,, \underline{q}) \big|_{x=x_0} = 0\,, \,\,\,\, \ell \geq 1\,, \, k \geq 0\,. \label{eq:psi1}
\ee
Again, the above conditions determine all $d_n$ and $\nabla_i$ derivatives of $\psi$ in the coincidence limit $x \rightarrow x_0$\,, and such a $\psi$ exists in a neighborhood of $x_0$ by Borel's theorem.

A covariant generalization of the Taylor series can be constructed. 
Since $\chi (x\,; x_0\,, \underline{\nu})$ is linear in $\nu$ and $\psi (x\,; x_0\,, \underline{q})$ is linear in $q_i$\,, we have
\be
    \chi (x\,; x_0\,, \underline{\nu}) = \nu \, \Phi^0 (x\,, x_0)\,,
        \qquad
    \psi (x\,; x_0\,, \underline{q}) = q_i \, \Phi^i (x\,, x_0)\,,
\ee
where $\Phi^\mu (x\,, x_0)$ defines a tangent vector. We then find, for a given function $f(x) \in C^\infty (\CM)$ and for $x$ in a neighborhood of $x_0$\,, 
\be
    f(x) = \sum_{k,\, \ell=0}^\infty \frac{1}{\ell!} \, d_n^k \, \nabla_{i_1} \cdots \nabla_{i_\ell} f (x_0) \, \bigl[ \Phi^0 (x\,, x_0) \bigr]^k \Phi^{i_1} (x\,, x_0) \cdots \Phi^{i_\ell} (x\,, x_0)\,.
\ee

We write the pseudodifferential operator $\CO$ using a covariant Fourier transform as
\be \label{eq:Omatel}
    \langle x | \CO | x_0 \rangle = \int \frac{d\nu}{2\pi} \frac{d^D\mathbf{q}}{(2\pi)^D\sqrt{h(x_0)}} \, e^{i \, \Phi(x;\,x_0,\,\{\underline{\nu}, \, \underline{q}\})} \, \sigma_\CO (x\,; x_0\,, \{\underline{\nu}, \, \underline{q}\})\,,
\ee
where $\sigma_\CO$ defines the symbol of $\CO$ and 
\be
    \Phi (x\,; x_0\,,\{ \underline{\nu}\,, \underline{q} \}) \equiv \chi (x\,; x_0\,, \underline{\nu}) + \psi (x\,; x_0\,, \underline{q})\,,
\ee
which reduces to the phase factor $\nu (t-t_0) + q_i (x^i-x^i_0)$ in the flat limit. Note that the frequency integral in \eqref{eq:Omatel} is over $\nu$, not the coefficient of $dt$ in $\underline{\nu}$, which is $N \nu$, as per the discussion immediately following \eqref{eq:nubar}. Had it been the latter, then the frequency integral in \eqref{eq:Omatel} would have had the customary factor of $N(x_0)$ in the denominator. 

The matrix element of the operator $\CO$ resolvent is
\begin{align} \label{eq:Ganiso}
    G (x\,, x_0\,|\,\lambda) & \equiv \langle x | (\CO - \lambda)^{-1} | x_0 \rangle \notag \\[2pt]
        & = \int \frac{d\nu}{2\pi} \frac{d^D\mathbf{q}}{(2\pi)^D\sqrt{h(x_0)}} \, e^{i \, \Phi(x;\,x_0,\,\{\underline{\nu}, \, \underline{q}\})} \, \sigma (x\,; x_0\,, \{\underline{\nu}, \, \underline{q}\} \,|\, \lambda)\,,
\end{align}
where $\sigma (x\,; x_0\,, \{\underline{\nu}, \, \underline{q}\} \,|\, \lambda) \equiv \sigma_{(\CO-\lambda)^{-1}} (x\,; x_0\,, \{\underline{\nu}, \, \underline{q}\})$\,. As in \eqref{eq:identity2}, \eqref{eq:Ganiso} implies that
\be
    \bigl[ \CO (x\,; d_n\,, \nabla) - \lambda \bigr] \, G (x, x_0\,|\,\lambda) = \frac{1}{N(x_0) \sqrt{h(x_0)}} \, \delta^{(D+1)} (x-x_0)\,. 
\ee
Therefore,
\be \label{eq:OlsI}
    \bigl[ \CO(x\,; d_n + i d_n \Phi\,, \nabla+i\nabla\psi) - \lambda \bigr] \, \sigma (x\,; x_0\,, \{ \underline{\nu}, \underline{q} \}) = I (x\,, x_0)\,,
\ee
where $I(x\,, x_0)$ satisfies
\be \label{eq:Idef2}
    \int \frac{d\nu}{2\pi} \frac{d^D \mathbf{q}}{(2\pi)^D} \, e^{i \Phi (x; \, x_0, \, \{\underline{\nu}, \,\underline{q}\})} \, I (x\,, x_0) = \frac{1}{N(x_0)} \, \delta^{(D+1)} (x - x_0)\,.
\ee
Again, had the integral been over the coefficient of $dt$ in $\underline{\nu}$, which is $N \nu$ as per \eqref{eq:nubar}, then the right hand side would just be the delta function. However, because we are integrating over $\nu$ and not $N \nu$, there is a factor of $N^{-1}$ on the right hand side. Using this convention, we can avoid carrying around factors of $N$ through the actual calculations in the following sections.

Using the conditions on the phase functions $\chi$ and $\psi$ introduced earlier in this subsection, we find from $\eqref{eq:Idef2}$ that 
\be \label{eq:Ico}
    I_{AB} (x_0\,, x_0) = \mathbb{1}_{AB}\,,
\ee
and
\begin{align} 
    \bigl( d_n \bigr){}_{A_1}{}^{\!B_1} \cdots \bigl( d_n \bigr){}_{A_\ell}{}^{\!B_\ell} \bigl( \nabla_{(i_{1}} \bigr){}_{C_{1}}{}^{\!D_{1}} \cdots \bigl(\nabla_{i_k)}\bigr){}_{C_k}{}^{\!D_k} I_{D_k E} (x\,, x_0) \, \big|_{x=x_0} & = 0\,, \label{eq:I1}
\end{align}
where $\ell + k \geq 1$\,. We recovered the bundle indices in $I$ and the covariant derivatives. For the scalar case without internal gauge symmetries, we simply have $I(x_0\,, x_0) = 1$\,. The above conditions on $I$ determine all derivatives of $I$\,.

\subsection{Heat Kernel Coefficients for Anisotropic Operators}

Finally, we discuss how to compute the heat kernel
\be \label{eq:KOaniso}
    \CK_\CO (x\,, x_0\,|\,\tau) = \langle x | e^{-\tau \CO} | x_0 \rangle = i \int_C \frac{d\lambda}{2\pi} \, e^{-\tau \lambda} \, G(x\,,x_0\,|\,\lambda)
\ee
for an 
operator $\CO$ in the coincidence limit $x_0 = x$\,, defined over a foliated spacetime.
Here, $C$ is a contour that bounds the spectrum of the operator $\CO$ in the complex plane and is traversed in the counter-clockwise direction.
In the following, we require that $\CO$ have an anisotropic scaling exponent $z$ at high energies, i.e. the UV fixed point is invariant under the rescaling of the spacetime coordinates
\be
    t \rightarrow b^{-z} \, t\,,
        \qquad
    \mathbf{x} \rightarrow b^{-1} \, \mathbf{x}\,.
\ee
Typically, we consider $\CO$ taking the form
\be \label{eq:Osk}
    \CO = - d_n^2 - (-1)^z \, \zeta^2 \, \nabla^{2z} + \cdots\,,
\ee
where we omitted the index structure of the spatial covariant derivatives and we note that there can be different terms that involve the same number of $\nabla$'s. Moreover, $``\cdots"$ denotes terms that contain fewer derivatives. We define the order of $\CO$ to be the dimension of its highest order term measured in the dimension of $\nabla$\,. For example, the order of $\CO$ in \eqref{eq:Osk} is $\Delta=2z$\,.

We start with expanding the symbol $\sigma \bigl(x\,; x_0\,, \{\underline{\nu}\,, \underline{q}\}\,\big| \lambda\bigr)$ as
\be
    \sigma \bigl(x\,; x_0\,, \{\underline{\nu}\,, \underline{q}\} \, \big|\, \lambda\bigr) = \sum_{m=0}^\infty \sigma^{(m)} \bigl(x\,; x_0\,, \{ \underline{\nu}\,, \underline{q}\} \, \big|\, \lambda\bigr)\,,
\ee
where $\sigma^{(m)}$ is a homogeneous function of $\lambda$\,, $\nu$\,, and $q_i$ in the coincidence limit, satisfying
\be \label{eq:sigmamaniso}
    \sigma^{(m)} \bigl(x_0\,; x_0\,, \{ b^z \, \underline{\nu}\,, b \, \underline{q}\} \, \big| \, b^{\Delta} \lambda \bigr) = b^{- m - \Delta} \sigma^{(m)} \bigl(x_0\,; x_0\,, \{ \underline{\nu}\,, \underline{q}\} \, \big| \, \lambda\bigr)\,.
\ee
This motivates us to consider the rescalings $\lambda \rightarrow b^\Delta \, \lambda$\,, $\nu \rightarrow b^z \nu$\,, $q_i \rightarrow b \, q_i$\,, and $\sigma^{(m)} \rightarrow b^{-m - \Delta} \sigma^{(m)}$ in \eqref{eq:OlsI}, supplemented with $\chi \rightarrow b^z \chi$ and $\psi \rightarrow b \, \psi$\,. We find
\be \label{eq:sumbI}
    \sum_{m=0}^\infty b^{-m-\Delta} \, \scD_b \, \sigma^{(m)} (x\,; x_0\,, \{ \underline{\nu}\,, \underline{q} \}\, | \, \lambda) = I (x\,, x_0)\,,
\ee
where 
\be
    \scD_b \equiv \CO \bigl( x\,; d_n + i \, b^z d_n \chi + i \, b \, d_n \psi\,, \nabla + i \, b \, \nabla \psi \bigr) - b^\Delta \lambda\,.
\ee
The above expansion with respect to $b$ is taken such that in the coincidence limit the relation \eqref{eq:sigmamaniso} is recovered.
Expand $\scD_b$ with respect to $b$\,, such that
\be \label{eq:Dbexp}
    \scD_b = \sum_{\ell=0}^\Delta b^{\Delta-\ell} \scD^{(\ell)} \bigl( x\,; d_n\,, \nabla\bigr)\,.
\ee
Plugging \eqref{eq:Dbexp} back into \eqref{eq:sumbI}, we find
\be \label{eq:rr}
    \sum_{m=0}^\infty \sum_{\ell=0}^\Delta b^{-m-\ell} \scD^{(\ell)} \sigma^{(m)} = I\,.
\ee
Matching the order in $b$ on both sides of \eqref{eq:rr} gives rise to a series of recursion relations that can be used to determine $\sigma^{(m)}$ in the coincidence limit. We will demonstrate how this works in explicit detail when we apply this method to the case of a Lifshitz operator on a foliated spacetime (see, e.g., \eqref{eq:Ds}).

Next, plugging \eqref{eq:Ganiso} into \eqref{eq:KOaniso}, and taking the rescalings 
\be
    \lambda \rightarrow \tau^{-1} \lambda\,, 
        \qquad
    \nu \rightarrow \tau^{-z/\Delta} \nu\,, 
        \qquad
    q_i \rightarrow \tau^{-1/\Delta} \, q_i\,, 
\ee
we find
\be \label{eq:asaniso}
    \CK_\CO (x_0\,, x_0 \, | \, \tau) = \sum_{m=0}^\infty \SG^{(m)} (x_0) \, \tau^{\frac{m - (D+z)}{\Delta}}\,,
\ee
where $\SG^{(m)} (x_0)$ are the heat kernel coefficients generalized to anisotropic operators, which are given by
\be \label{eq:amnoniso}
    \SG^{(m)} (x_0) = \int \frac{d\nu}{2\pi} \, \frac{d^D \mathbf{q}}{(2\pi)^{D} \sqrt{h (x_0)}} \int_C \frac{i \, d\lambda}{2\pi} \, e^{-\lambda} \, \sigma^{(m)} \bigl(x_0\,; x_0\,, \{\underline{\nu}\,, \underline{q} \}\, \big|\, \lambda \bigr)\,.
\ee
Again, if $m$ is odd, then either the number of frequency factors or the number of momentum factors in the integrand of $\SG^{(m)}$ is odd. Thus, $\SG^{(m)} = 0$ when $m$ is odd and the heat kernel can be written as
\be \label{eq:hkfinal}
    \CK_\CO (x_0\,, x_0 \, | \, \tau) = \sum_{r=0}^\infty \SG^{(2r)} (x_0) \, \tau^{\frac{2r - (D+z)}{\Delta}}\,.
\ee
Note that, when $z = 1$\,, the asymptotic expansion \eqref{eq:asaniso} around $\tau \rightarrow 0^+$ reduces to \eqref{eq:Seeley}. This concludes our formal discussion on the covariant heat kernel method for operators that involve both $d_n$ and $\nabla$.

\section{Lifshitz Scalar Field Theories in \texorpdfstring{$2+1$}{2+1} Dimensions} \label{sec:examples}

In this section, we will use the simplest examples of scalars on an anisotropic gravitational background in $2+1$ dimensions to illustrate how the covariant heat kernel method proceeds in practice. However, the method developed in this paper is applicable to general dimensions as well as gauge vector fields and more general tensor fields.

\subsection{Anisotropic Weyl Anomaly for Lifshitz Scalar} \label{sec:awals}

We first consider a single real scalar field at a $z=2$ Lifshitz fixed point in $2+1$ dimensions. We take the time to be imaginary. We focus on the following action:
\begin{align} \label{eq:Weyl}
    S & = \frac{1}{2} \int dt \, d^2 \mathbf{x} \, N \sqrt{h} \, \Bigl[ \bigl( d_n \phi \bigr)^2 + \bigl( \Box \, \phi \bigr)^2 \Bigr] \notag\\[2pt]
    & = \frac{1}{2} \int dt \, d^2 \mathbf{x} \, N \sqrt{h} \, \phi \, \CO \phi\,,
\end{align}
where $\Box \equiv \nabla_i \nabla^i$ and
\begin{align} \label{eq:Ononp}
    \CO (x\,; d_n\,, \nabla) & = - d_n^2 - K \, d_n + \frac{1}{N(x)} \, \Box \, N(x) \, \Box \notag \\[2pt]
    & = - d_n^2 - K \, d_n + \Box^2 - 2 \, a^i \, \nabla_{\!i} \, \Box + \bigl( a^i \, a_i - \nabla^i a_i \bigr) \, \Box\,. 
\end{align}
Here, $x = (t\,, \mathbf{x})$\,. As we remarked previously, the operator $\CO$ can depend on the fields and their derivatives. Here, for example, $\CO$ depends on time and spatial derivatives of the spatial metric, lapse, and shift, in the form of the extrinsic curvature $K$, acceleration vector $a_i$, and the derivative of the acceleration vector, $\nabla^i a_i$. The derivatives in those expressions do \emph{not} continue on to act on the scalar field $\phi$. Therefore, when the latter is covariantly Fourier-transformed, only the derivatives that actually act on $\phi$ get shifted, not the derivatives acting on the background fields.

Take the engineering dimensions for space and time as
\be
    [t] = -2\,,
        \qquad
    [x^i] = - 1\,.
\ee
The scaling dimensions of the derivatives, background, and scalar fields, and the operator $\CO$ then follow:
\be
    [d_n] = 2\,,
        \qquad
    [\nabla_i] = 1\,,
        \qquad
    [N] = [N_i] = [h_{ij}] = [\phi] = 0\,,
        \qquad
    \Delta = [\CO] = 4\,.
\ee

Note that the action \eqref{eq:Weyl} is classically invariant under the local anisotropic Weyl transformation,
\be
    N \rightarrow e^{-z \, \Omega(x)} \, N\,,
        \qquad
    N_i \rightarrow e^{-2 \, \Omega(x)} \, N_i\,,
        \qquad
    h_{ij} \rightarrow e^{-2 \, \Omega(x)} \, h_{ij}\,,
        \qquad
    \phi \rightarrow \phi\,.
\ee
The infinitesimal anisotropic Weyl transformation of the effective action $\Gamma$ as introduced in \eqref{eq:effaction} can be anomalous, with
\be
    \Omega (x) \lr z \, N \, \frac{\delta}{\delta N} + 2 \, N_i \frac{\delta}{\delta N_i} + 2 \, h_{ij} \frac{\delta}{\delta h_{ij}} \rr \Gamma (x) = \CA (x)\,,
\ee
where $\CA$ denotes the Weyl anomaly \cite{Griffin:2011xs}. To determine the form of the anomaly, we introduce a ghost field $c$ and the nilpotent BRST operator
\be
    s = c \lr z \, N \, \frac{\delta}{\delta N} + 2 \, N_i \frac{\delta}{\delta N_i} + 2 \, h_{ij} \frac{\delta}{\delta h_{ij}} \rr,
\ee
which acts on $c$ trivially. Then, the Wess-Zumino consistency condition \cite{Weinberg:1996kr} requires that
\be
    s \int dt \, d^2 \mathbf{x} \, N \sqrt{h} \, \CA \, c = 0\,.
\ee
If an anomaly term is BRST exact, i.e., this term can be expressed as a BRST variation of a local operator, then it can be subtracted from the action, thus canceling the associated anomaly. As a result, the possible anomalies of interest are those in the cohomology of the BRST differential $s$\,. It turns out that all terms in this cohomology group are linear combinations of the following two terms \cite{Baggio:2011ha, Griffin:2011xs}:
\be \label{eq:anom}
    K^{ij} K_{ij} - \frac{1}{2} K^2\,,
        \qquad
    \lr R + \nabla^i a_i \rr^2,
\ee
which are of dimension $\Delta = 4$\,, measured in spatial momentum. In higher than $(2+1)$-dimensions, other terms can show up depending on the Riemann or Ricci tensors \cite{Arav:2014goa}, but in the case of a two-dimensional leaf, the Riemann and Ricci tensors are related to the Ricci scalar as
\begin{align} \label{eq:riemann2d}
    R_{ijk\ell} = \frac{1}{2} \, R \, \bigl( h_{ik} \, h_{j\ell} - h_{i\ell} \, h_{jk} \bigr) \,, \qquad %
    R_{ij} = \frac{1}{2} \, R \, h_{ij} \,.
\end{align}
In \cite{Arav:2014goa, Baggio:2011ha}, it is shown that there is one time-derivative term that is BRST exact,
\be \label{eq:KKdK}
    K^2 + d_n K = \frac{1}{N \sqrt{h}} \Bigl[ \p_t \bigl( \sqrt{h} \, K \bigr) - \p_i \bigl( \sqrt{h} \, N^i \, K \bigr) \Bigr]\,.
\ee
This is a total derivative term.
It has also been shown in \cite{Griffin:2011xs, Baggio:2011ha, Arav:2014goa} that there are five independent BRST exact terms with only spatial derivatives, which are total derivatives that take the form
\be \label{eq:exactterms}
    \frac{1}{N \sqrt{h}} \, \p_i \bigl( \sqrt{h} \, F^i_I \bigr)\,,
        \qquad
    I = 1\,, \cdots, 5\,,
\ee
where, in the basis chosen in \cite{Griffin:2011xs}, 
\begin{subequations} 
\begin{align}
    & F^i_1 = \nabla^i \bigl[ N (R - \nabla_{\!i} \, a^i) \bigr] \,, \\[2pt]
    & F^i_2 = - N (R - \nabla^j a_j) \, a^i\,, \\[2pt]
    & F^i_3 = - \nabla^i \bigl( N \, \nabla_{\!j} \, a^j \bigr)\,, \\[2pt]
    & F^i_4 = N \, (\tfrac{1}{2} \, a^2 + \nabla_{\!j} \, a^j - a^j \, \nabla_{\!j}) \, a^i\,, \\[2pt]
    & F^i_5 = - N \, a^2 \, a^i\,. 
\end{align}
\end{subequations}
We are interested in the heat kernel coefficient associated with the anisotropic Weyl anomalies \eqref{eq:anom} and the total derivative terms  \eqref{eq:KKdK} and \eqref{eq:exactterms}, and reproducing the result in \cite{Griffin:2011xs, Baggio:2011ha} using the new method developed here. 
All such terms have scaling dimension four and are thus marginal. Therefore, the heat kernel coefficient of interest is $\SG^{(4)}$. In fact, it is shown in \cite{Baggio:2011ha} that 
$\CA (x) = - 2 \, \SG^{(4)} (x)$\,. 

In the following, we compute $\SG^{(4)}$ using the covariant heat kernel method. We start by plugging \eqref{eq:Ononp} into \eqref{eq:Dbexp}, and setting $z=2$ and $\Delta = 4$\,, which gives
\begin{align}
    \scD_b
    & = \CO (x\,; d_n + i \, b^z \, d_n \chi + i \, b \, d_n \psi\,, \nabla+ i \, b \, \nabla \psi) - b^\Delta \lambda \notag \\[2pt]
    & = - (d_n + i \, b^2 \, d_n \chi + i \, b \, d_n \psi)^2 - K \, (d_n + i \, b^2 \, d_n \chi + i \, b \, d_n \psi) \notag \\[2pt]
    & \quad + \bigl[ \bigl( \nabla_i + i \, b \, \psi_i \bigr) \bigl( \nabla^i + i \, b \, \psi^i \bigr) \bigr]^2 \notag \\[2pt]
    & \quad - \ls 2 \, a^i \bigl( \nabla_{\!i} + i \, b \, \psi_i \bigr) - a^i \, a_i + \nabla^i a_i \rs \bigl( \nabla_{\!j} + i \, b \, \psi_j \bigr) \bigl( \nabla^j + i \, b \, \psi^j \bigr)  - b^4 \lambda \notag \\[2pt]
    & = \sum_{\ell = 0}^4 b^{4-\ell} \scD^{(\ell)} (x\,; d_n\,, \nabla)\,,
\end{align}
where
\begin{subequations} \label{eq:Dmnonp}
\begin{align}
	\scD^{(0)} & = \lr d_n \chi \rr^2 + \lr \psi_i \psi^i \rr^2 - \lambda\,, \\[5pt]
	\scD^{(1)} & = 2 \, d_n \chi \, d_n \psi - 2 \, i \Bigl[ 2 \, \psi^i \psi^j \psi_{ij} + \psi^i \psi_i \lr 2 \, \psi^j \nabla_{\!j} + \Box \psi \rr \Bigr] + 2 \, i \, a^i \, \psi_i \, \psi_j \, \psi^j, \\[5pt]
	\scD^{(2)} & = - 2 \, i \, d_n \chi \, d_n - 2 \Bigl[ 2 \, \psi^i \, \psi^j \, \nabla_{\!i} \nabla_{\!j} + \psi_i \, \psi^i \, \Box + 2 \lr \psi^i \psi_j{}^j + 2 \, \psi_j \, \psi^{ij}\rr \nabla_{\!i} \Bigr] - \bigl( \psi_i{}^i \bigr)^2 \notag \\[2pt]
	& \quad - i \bigl( d_n^2 \chi + K d_n \chi + i \, d_n \psi \, d_n \psi \bigr) - 2 \Bigl[ \psi_{ij} \psi^{ij} + \psi^i \lr \psi^j{}_{ji} + \psi_{ij}{}^j \rr \Bigr] \notag \\[2pt]
	& \quad + 2 \, a^i \Bigl[ \psi_i \, \psi^j{}_j + \psi^j \, \bigl( 2 \, \psi_{ij} + 2 \, \psi_i \nabla_{\!j} + \psi_j \nabla_{\!i} \bigr) \Bigr] + \bigl( \nabla_{\!j}\,a^j - a^2 \bigr) \, \psi_i \, \psi^i\,, \\[5pt]
	\scD^{(3)} & = - i \, \bigl( 2 \, d_n \psi \, d_n + d^2_n \psi + K \, d_n \psi \bigr) + i \, \psi_i{}^i{}_j{}^j \notag \\[2pt]
	& \quad + 2 \, i \bigl[ \psi^i \lr \nabla_{\!i} \Box + \Box \nabla_{\!i} \rr + \lr \psi^j{}_{ji} + \psi_{ij}{}^j \rr \nabla^i + \psi_i{}^i \, \Box + 2 \, \psi^{ij} \, \nabla_{\!i} \nabla_{\!j} \bigr] \notag \\[2pt]
	& \quad - 2 \, i \, a^i \bigl( \psi_i{}^j{}_j + \psi^j{}_j \, \nabla_i + 2 \, \psi^j \, \nabla_i \nabla_j + 2 \, \psi_i{}^j \, \nabla_{\!j} + \psi_i \, \Box \bigr) \notag \\[2pt]
	& \quad - i \, \bigl( \nabla_{\!j} \, a^j - a^2 \bigr) \bigl( \psi^i{}_i + 2 \, \psi^i \, \nabla_i \bigr)\,, \\[5pt]
	\scD^{(4)} & = - \lr d_n + K \rr d_n+ \Box^2 - \bigl( 2 \, a^i \, \nabla_{\!i} - a^i \, a_i + \nabla^i a_i \bigr) \, \Box \,.
\end{align}
\end{subequations}
We introduced the notation $\psi_{i_1 \cdots i_k} \equiv \nabla_{\!i_1} \cdots \nabla_{\!i_k} \psi$\,.\footnote{Covariant derivatives are sometimes denoted in the literature by a semicolon preceding the indices of the derivatives. In this case, the order of the indices reads left-to-right the order in which the covariant derivatives actually \emph{act}, not the order in which they would be written left-to-right: $\nabla_{i_1} \cdots \nabla_{i_k} \psi = \psi_{;i_k \cdots i_1}$. This is, for example, the convention used in \cite{Gilkey:1975iq}.}

Plugging \eqref{eq:Dmnonp} into \eqref{eq:rr}, we find the following recursion relations:
\begin{subequations} \label{eq:Ds}
\begin{align}
    \scD^{(0)} \, \sigma^{(0)} & = I\,, \label{eq:Ds0} \\[2pt]
    \scD^{(0)} \, \sigma^{(1)} + \scD^{(1)} \, \sigma^{(0)} & = 0\,, \label{eq:Ds1} \\[2pt]
    \scD^{(0)} \, \sigma^{(2)} + \scD^{(1)} \, \sigma^{(1)} + \scD^{(2)} \, \sigma^{(0)} & = 0\,, \label{eq:Ds2} \\[2pt]
    \scD^{(0)} \, \sigma^{(3)} + \scD^{(1)} \, \sigma^{(2)} + \scD^{(2)} \, \sigma^{(1)} + \scD^{(3)} \, \sigma^{(0)} & = 0\,, \\[2pt]
    \scD^{(0)} \, \sigma^{(k)} + \scD^{(1)} \, \sigma^{(k-1)} + \scD^{(2)} \, \sigma^{(k-2)} + \scD^{(3)} \, \sigma^{(k-3)} + \scD^{(4)} \, \sigma^{(k-4)} & = 0\,, 
\end{align}
\end{subequations}
for $k \geq 4$. From this set of recursion relations, we solve for $\sigma^{(0)}, \cdots, \sigma^{(4)}$ and calculate the associated heat kernel coefficients. Clearly, to calculate $\sigma^{(m)}$, we must also calculate various derivatives acting on $\sigma^{(0)}$ up to $\sigma^{(m-1)}$. Generically, we will need up to $m$ derivatives on $\sigma^{(0)}$, $m-1$ on $\sigma^{(1)}$, and so on up to one derivative on $\sigma^{(m-1)}$ (with time derivatives counting as $z$ spatial derivatives). To compute these, we simply take derivatives of the appropriate recursion relation. For example, to calculate $\sigma^{(1)}$, we will need $\nabla_{\!i}\, \sigma^{(0)}$, which we compute by taking $\nabla_{\!i}$ of \eqref{eq:Ds0}. We detail this procedure as follows:

\vspace{3mm}

\noindent \textbf{Coincidence Limit of $\sigma^{(0)}$\,.} From \eqref{eq:Ds0}, using the coincidence limits 
of derivatives of $\chi$\,, $\psi$ and $I$ given in \S\ref{sec:cftao}, namely,
\be
    d_n \chi \big|_{x=x_0} = \nu\,,
        \qquad
    \nabla_{\!i} \psi \big|_{x=x_0} = q_i\,,
        \qquad
    I \big|_{x=x_0} = 1\,,
\ee
we find 
\be
    \sigma^{(0)} (x_0\,; x_0\,, \{\underline{\nu}\,, \underline{q}\} \, | \, \lambda) = \frac{1}{\nu^2 + |\mathbf{q}|^4 - \lambda} \equiv \CLG\,.
\ee
This is essentially the propagator of the theory.

From \eqref{eq:amnoniso}, we find that the heat kernel coefficient $\SG^{(0)}$ is given by
\begin{align} \label{eq:a0exp}
    \SG^{(0)} (x_0) = \int \frac{d\nu}{2 \pi} \frac{d^2 \mathbf{q}}{(2\pi)^{2} \sqrt{h (x_0)}} \int_C \frac{i \, d\lambda}{2\pi} \, e^{- \lambda} \, \sigma^{(0)} \bigl(x_0\,; x_0\,, \{\underline{\nu}\,, \underline{q} \}\, \big|\, \lambda \bigr) = \frac{1}{16\,\pi}\,.
\end{align}

\vspace{3mm}

\noindent \textbf{Coincidence Limit of $\sigma^{(1)}$\,.} Since $\psi$ is a scalar, the two derivatives in $\nabla_{\!i} \nabla_{\!j} \, \psi$ commute. Therefore, the defining relation $\nabla_{\!(i} \nabla_{\!j)} \psi \big|_{x=x_0}$ in \eqref{eq:dpc} implies
\be
    \nabla_{\!i} \nabla_{\!j} \, \psi \big|_{x=x_0} = 0\,.
\ee
The defining relation \eqref{eq:psi1}, with the number of time derivatives set to $\ell = 1$ and the number of spatial derivatives set to $k=0$, reads
\be
    d_n \psi \big|_{x=x_0} = 0\,.
\ee
Then, from \eqref{eq:Ds1}, we obtain
\be \label{eq:ds1}
    \sigma^{(1)} \big|_{x=x_0} = - \CLG \,  \scD^{(1)} \sigma^{(0)} \big|_{x=x_0} = 4 \, i \, \CLG \, |\mathbf{q}|^2 \, q^i \, \nabla_{\!i} \, \sigma^{(0)} \big|_{x=x_0}- 2 \, i \, \CLG^2 \, a^i \, q_i \, |\mathbf{q}|^2\,.
\ee
To derive the coincidence limit of $\nabla_{\!i} \, \sigma^{(0)}$ in \eqref{eq:ds1},  we act $\nabla_{\!i}$ on both sides of \eqref{eq:Ds0} and then take the coincidence limit. When $\nabla_{\!i}$ acts on the phase functions in $\scD^{(0)}$, it will produce factors involving $\nabla_{\!i} \, d_n \chi$\,, whose coincidence limit can be derived from \eqref{eq:chi1} by commuting $\nabla_{\!i}$ past $d_n$ using \eqref{eq:dnnabcomm}. The result is
\be
    \nabla_{\!i} \, d_n \chi \big|_{x=x_0} = \nu \, a_i\,.
\ee
Moreover, $\nabla_{\!i} I \bigr|_{x=x_0} = 0$ by \eqref{eq:I1}, so the right hand side vanishes. In the end, we find
\be \label{eq:ds0result}
    \nabla_{\!i} \, \sigma^{(0)} \big|_{x=x_0} = - 2 \, \CLG^2 \, \nu^2 \, a_i\,.
\ee
Plugging \eqref{eq:ds0result} back into \eqref{eq:ds1} gives 
\be
    \sigma^{(1)} \big|_{x=x_0} = - 2 \, i \, \CLG^2 \, a^i \, q_i \, |\mathbf{q}|^2 \, \bigl( 1 + 4 \, \CLG \, \nu^2 \bigr)\,. 
\ee
As expected, $\sigma^{(1)}$ is odd in $q_i$ and will vanish upon integration over momentum. Thus, the associated heat kernel coefficient vanishes:
\begin{equation}
    \SG^{(1)} (x_0) = 0\,.
\end{equation}

\vspace{3mm}

\noindent \textbf{Coincidence Limit of $\sigma^{(2)}$\,.} 
From the conditions given in \S\ref{sec:cftao},
\be
    d_n \nabla_{\!i} \, \nabla_{\!j} \chi = 
    d_n \nabla_{\!} \psi \big|_{x=x_0} =
    \psi_{(ijk)} \big|_{x=x_0} = 0\,,
\ee
we find the following coincidence limits:
\begin{subequations}
\begin{align}
    \nabla_{\!i} \nabla_{\!j} \, d_n \chi \big|_{x=x_0} & = \nu \bigl( a_i \, a_j + \nabla_{\!i} \, a_j \bigr)\,, 
        \qquad
    \qquad
    \nabla_{\!i} \, d_n \psi \big|_{x=x_0} = 0\,, \\[2pt]
    \psi_{ijk} \big|_{x = x_0} & = \frac{1}{6} \, R \, \bigl( h_{ij} \, q_k + h_{ik} \, q_j - 2 \, q_i \, h_{jk} \bigr)\,,
\end{align}
\end{subequations}
Using \eqref{eq:Ds0} and \eqref{eq:Ds1} and the above coincidence limits of $\chi$ and $\psi$\,, we derive
\begin{align*}
    d_n \sigma^{(0)} \big|_{x=x_0} & = 4 \, \CLG^2  |\mathbf{q}|^2 \, q_i \, q_j \, K^{ij}, \\[2pt]
    \nabla_{\!i} \nabla_{\!j} \, \sigma^{(0)} \big|_{x=x_0} & = \frac{2}{3} \, \CLG^2 \, |\mathbf{q}|^2 \, \bigl( q_i \, q_j - |\mathbf{q}|^2 \, h_{ij} \bigr)\, R \notag \\[2pt]        
    & \quad - 2 \, \CLG^2 \, \nu^2 \Bigl[ \nabla_{\!i} \, a_j + 2\, a_i \, a_j \bigl( 1 - 2 \, \CLG \, \nu^2 \bigr) \Bigr]\,, \\[2pt]
    \nabla_{\!i} \, \sigma^{(1)} \big|_{x=x_0} & = - \frac{2 \, i}{3} \, \CLG^2 \, |\mathbf{q}|^2 \, q_i \, R \notag \\[2pt]
    & \quad - 2 \, i \, \CLG^2 \, |\mathbf{q}|^2 \, q^j \, \Bigl[ 4 \, a_i \, a_j \, \CLG \, \nu^2 \bigl( 1 - 6 \, \CLG \, \nu^2 \bigr) + \nabla_{\!i} \, a_j \bigl( 1 + 4 \, \CLG \, \nu^2 \bigr) \Bigr]\,.
\end{align*}
Finally, plugging the coincidence limits that we have derived so far into \eqref{eq:Ds2}, we find
\begin{align} \label{eq:s2weyl}
    \sigma^{(2)} \big|_{x=x_0} & = - \frac{1}{3} \, \CLG^2 \, |\mathbf{q}|^2 \, (1 - 4 \, \CLG \, |\mathbf{q}|^4) R + i \, \CLG^2 \, \nu \, \bigl( h_{ij} + 8 \, \CLG \, q_i \, q_j \, |\mathbf{q}|^2 \bigr)\, K^{ij} \notag \\[2pt]
    & \quad - \CLG^2 \, \nabla_{\!i} \, a^i \, |\mathbf{q}|^2 \, \bigl( 1 + 4 \, \CLG \, \nu^2 \bigr) + 8 \, \CLG^3 \, \nabla_{\!i} \, a_{j} \, q^i \, q^j \Bigl[ |\mathbf{q}|^4 - (1 - 4 \, \CLG \, |\mathbf{q}|^4) \, \nu^2 \Bigr] \notag \\[2pt]
    & \quad - 4 \, \CLG^3 \, (\mathbf{a} \cdot \mathbf{q})^2 \Bigl[ |\mathbf{q}|^4 + 2 \, \bigl( 1 - 2 \, \CLG \, |\mathbf{q}|^4 \bigr) \, \nu^2 - 8 \, \CLG (1 - 6 \, \CLG \, |\mathbf{q}|^4) \, \nu^4 \Bigr] \notag \\[2pt]
    & \quad + \CLG^2 \, a^2 \, |\mathbf{q}|^2 \, \Bigl[ 1 - 4 \, \CLG \, \nu^2 \bigl( 1 - 4 \, \CLG \, \nu^2 \bigr) \Bigr]\,.
\end{align}
We also give a step-by-step derivation of $\sigma^{(2)}$ in Appendix \ref{app:E2}. From \eqref{eq:amnoniso}, we obtain the heat kernel coefficient $\SG^{(2)}$\,,
\begin{align} \label{eq:a2exp}
    \SG^{(2)} (x_0) & =  \int \frac{d\nu}{2 \pi} \int \frac{d^2 \mathbf{q}}{(2\pi)^{2} \sqrt{h (x_0)}} \int_C \frac{i \, d\lambda}{2\pi} \, e^{- \lambda} \, \sigma^{(2)} \bigl(x_0\,; x_0\,, \{\underline{\nu}\,, \underline{q} \}\, \big|\, \lambda \bigr) \notag \\[2pt]
    & = \frac{1}{48 \, \pi^{3/2}} \, \bigl( R + \nabla_{\!i} \, a^i - a_i \, a^i \bigr)\,,
\end{align}
where we used the integral
\begin{align} \label{eq:integral}
   I_{z, \, D, \, j, \, k, \,\ell} & \equiv \int \frac{d\nu}{2\pi} \, \nu^{2j}  \int \frac{d^D \mathbf{q}}{(2\pi)^D \sqrt{h}} \, q_{i_1} \cdots q_{i_{2k}} \int_C \frac{i \, d\lambda}{2\pi} \, \frac{e^{-\lambda}}{(\nu^2 + |\mathbf{q}|^{2z} - \lambda)^\ell} \notag\\[2pt]
    & = \frac{2^{1-k}}{(4\pi)^{\frac{D}{2} +1}} \, \frac{1}{z} \, \frac{\Gamma\bigl(j+\tfrac{1}{2}\bigr) \, \Gamma \bigl( \tfrac{k}{z} +\tfrac{D}{2z} \bigr)}{\Gamma\bigl(k+\tfrac{D}{2}  \bigr) \, \Gamma \bigl(\ell\bigr)} 
    \, h_{i_1 \cdots i_{2k}}\,.
\end{align}
This is a straightforward integral, which is essentially Gaussian once the integral over $\lambda$ is performed using the residue theorem. It does require us to convert a product of momenta with free indices to a product of pairwise-contracted momenta. Provided that the rest of the integral is rotationally invariant in momentum space or, in other words, is a function only of $| \bq |^2$, then we can perform the following replacement:
\begin{equation}
    q_{i_1} \cdots q_{i_{2k}} \rightarrow \frac{\Gamma \bigl( \frac{D}{2} \bigr)}{2^k \, \Gamma \bigl( k + \frac{D}{2} \bigr)} \, q^{2k} h_{i_1 \cdots i_{2k}},
\end{equation}
where $h_{i_1 \cdots i_{2k}}$ is the symmetrized combination of $h_{ij}$'s. For example,
\be
    h_{ijk\ell} \equiv h_{ij} \, h_{k\ell} + h_{ik} \, h_{j\ell} + h_{i\ell} \, h_{jk}\,.
\ee
In the relativistic case, the appropriate integral is the same, but without the $\nu$ integral in \eqref{eq:integral} and setting $z=1$. 
Notice that setting $z=1$, the factor of $\Gamma \bigl( \frac{k}{z} + \frac{D}{2z} \bigr)$ in the numerator of \eqref{eq:integral} cancels the factor of $\Gamma \bigl( k + \frac{D}{2} \bigr)$ in the denominator and, besides the overall power of $4 \pi$, 
the dependence on $D$ completely drops out of this integral. This is why the heat kernel coefficients in the relativistic case do not explicitly depend on the dimension besides the overall power of $4 \pi$. For other values of $z$, however, the heat kernel coefficients will explicitly depend on the spacetime dimension in addition to the overall power of $4\pi$\,. Of course, we will not expose this extra dimension-dependence in this work because our calculations are performed specifically in $D=2$.

Looking back at $\SG^{(2)}$ in \eqref{eq:a2exp}, note that the combination $N \sqrt{h} \, ( \nabla_{\!i} \, a^i - a_i \, a^i )$ is a total derivative and, in two dimensions, the term $N \sqrt{h} \, R$ is also a total derivative. Therefore, $E^{(2)}$ vanishes once integrated over a spacetime with no boundaries. Our results for $\SG^{(0)}$ and $\SG^{(2)}$ agree with the ones in Appendix D of \cite{Baggio:2011ha}.

\vspace{3mm}

\noindent \textbf{Heat Kernel Coefficient $E^{(4)}$\,.} Similarly, a more involved process that we implemented using xAct \cite{xAct} on Mathematica leads to the results $\SG^{(3)} = 0$ and  
\begin{align} \label{eq:a4re}
    \SG^{(4)} 
    & = - \frac{1}{64 \pi} \lr K_{ij} K^{ij} - \frac{1}{2} \, K^2 \rr - \frac{1}{48 \pi} \bigl( d_n K + K^2 \bigr) + \frac{1}{N \sqrt{h}} \, \p_i f^i\,, 
\end{align}
where
\begin{align}
    f^i & = \frac{1}{960 \pi} \sqrt{h} \, \Bigl\{ N \, \bigl[ 5 \, (R + 2 \,\Box) - 4 \, \nabla_{\!j} \, a^j - 16 \, a^j \nabla_{\!j} \bigr] \, a^i \Bigr\} \notag \\[2pt]
    & = - \frac{1}{480} \, \sqrt{h} \, \bigl( 5 \, F_2^i + 5 \, F_3^i - 8 \, F_4^i - 4 \, F_5^i \bigr) \,,
\end{align}
and $F_I^i$\,, $I = 2, \cdots, 5$ are defined in \eqref{eq:exactterms}. We used the identity 
\be
    \nabla_{\!i} \, a_j = - \nabla_{\!i} \, \nabla_{\!j} \, \ln N = \nabla_{\!j} \, a_i\,.
\ee
The combinations $K_{ij} \, K^{ij} - \tfrac{1}{2} \, K^2$ and $d_n K + K^2$ are BRST-invariant terms classified in \eqref{eq:anom} and \eqref{eq:KKdK}. In particular, as indicated in \eqref{eq:KKdK}, $d_n K + K^2$ is a total derivative term. Therefore, $E^{(4)}$ in \eqref{eq:a4re} is a linear combination of the BRST-invariant terms in \eqref{eq:anom} $\sim$ \eqref{eq:exactterms}, exactly reproducing the result in \cite{Baggio:2011ha, Griffin:2011xs}. This matching provides us with a rather strong check of our method.

\vspace{3mm}

The contribution to the diagonal heat kernel in \eqref{eq:hkfinal} is
\be
    \CK_\CO (x_0\,, x_0 \, | \, \tau) = \SG^{(0)} (x_0) \, \tau^{- 1} + \SG^{(2)} (x_0) \, \tau^{- 1/2} + \SG^{(4)} (x_0) + O (\tau^{1/2})\,.
\ee
While $\SG^{(0)}$ and $\SG^{(2)}$ contribute power law divergences to the one-loop effective action \eqref{eq:hkrep} that can be set to zero by adding in appropriate counterterms, $\SG^{(4)}$ contributes a log divergence. The one-loop effective action defined in \eqref{eq:hkrep} gives
\begin{align} \label{eq:Gamma1W}
    \Gamma_1 & = - \frac{1}{2} \frac{d}{ds} \Big|_{s=0} \frac{\mu^{2s}}{\Gamma(s)} \int dt \, d^D \mathbf{x} \, N \sqrt{h} \int_{0}^{1/M^2} d\tau \, \tau^{s-1} \, \SG^{(4)} (t, \mathbf{x}) + \text{finite} \notag \\[2pt]
    & = - \frac{1}{128 \pi} \lr \log \frac{M^2}{\mu^2} - \gamma^{}_E \rr \int dt \, d^2 \mathbf{x} \, N \sqrt{h} \, \lr K_{ij} K^{ij} - \frac{1}{2} \, K^2 \rr 
    + \text{finite}\,,
\end{align}
where $\gamma_E$ is the Euler-Mascheroni constant. We have introduced a cutoff for the $\tau$-integral, with $M$ effectively acting as a UV cutoff, while $\mu$ acts as an infrared regulator. 
The theory exhibits an anomaly under the local anisotropic scale transformation (in momentum), 
\be
    \CA = \frac{1}{32 \pi} \lr K_{ij} K^{ij} - \frac{1}{2} \, K^2 \rr. 
\ee
The fact that the second term in \eqref{eq:anom} does not show up in the anomaly at this one-loop order can be explained by the detailed balance condition \cite{Griffin:2011xs}.
This corroborates previous results which were computed using the (non-covariant) plane wave method for evaluating the heat kernel \cite{Baggio:2011ha} and also from the holographic renormalization calculation \cite{Griffin:2011xs, Baggio:2011ha, Griffin:2012qx}. In contrast to the previous heat kernel calculation in \cite{Baggio:2011ha}, however, our method is fully covariant and does not assume any special ansatz for the spacetime metric, which may prove useful for a systematic study of more complicated scenarios.

Note that $\SG^{(2)}$ contributes a power-law divergence to the effective action. The Wess-Zumino consistency condition then requires $\SG^{(2)}$ to be zero up to total derivatives. There are four terms that can show up in $E^{(2)}$\,: $K$, $R$, $a_i \, a^i$ and $\nabla_{\!i} \, a^i$, which all have scaling dimension 2. It will be convenient to choose $K$, $R$, $\nabla_{\!i} \, a^i - a_i \, a^i$ and $a_i \, a^i$ as the basis elements instead. The reason for this is that $NK$ is a total time derivative and the combination $N \sqrt{h} \, ( \nabla_{\!i} \, a^i - a_i \, a^i )$ is a total space derivative and can be dropped. 
In two spatial dimensions, $N \sqrt{h} \, R$ also happens to be a total derivative and can be dropped as well. 
The only coefficient left is that of $a_i \, a^i$, which is forced to vanish by the Wess-Zumino consistency condition (once part of it is combined with $\nabla_{\!i} \, a^i$ in the form of $\nabla_{\!i} \, a^i - a_i \, a^i$). Indeed, our result for $\SG^{(2)}$ in \eqref{eq:a2exp} bears this out. 

Table 1 in \cite{DOdorico:2015pil} reports some heat kernel coefficients for Lifshitz scalars with $z=2$ and $z=3$ in $D=2$ and $D=3$\,. In particular, for $z=D=2$, they report that the coefficient of $a_i a^i$ is $- \frac{13}{12}$\,. However, the operator studied in \cite{DOdorico:2015pil} is \emph{not} the same as the one we have just studied here and which was studied previously in \cite{Baggio:2011ha, Griffin:2011xs}. The spatial derivative part of the classically Weyl-invariant operator is $N^{-1} \Box N \Box$. In contrast, the operator studied in \cite{DOdorico:2015pil} is $\Delta_{x}^{z}$\,, where 
\be
    \Delta_x = - \frac{1}{N \sqrt{h}} \partial_i N \sqrt{h} \, h^{ij} \partial_j = N^{-1} \nabla^i N \nabla_i\,.
\ee
For $z=2$, the operator $\Delta_x^2$ is equal to $N^{-1} \nabla^i N \nabla_i N^{-1} \nabla^j N \nabla_j$\,, which is not the same as $N^{-1} \Box N \Box$. However, in the next subsection, we will study the most general $z=2$ scalar operator in $2+1$ dimensions and this will obviously include the one considered in \cite{DOdorico:2015pil}. Thus, we will return to the question of their $a_i \, a^i$ coefficient at the end of the next subsection.

\subsection{General Scalar Operators Around a \texorpdfstring{$z=2$}{z=2} Lifshitz Point}

We have tested our method in the last subsection for a well-known example. Now, we take one step forward and consider the most general scalar operator in $2+1$ dimensions, around a $z=2$ Lifshitz fixed point.\footnote{However, we will impose the condition that the operator does not mix time and space derivatives.} The associated action principle is 
\begin{align} \label{eq:genaction}
    S & = \frac{1}{2} \int dt \, d^2 \mathbf{x} \, N \sqrt{h} \, \phi \, \tilde{\mathcal{O}} \, \phi\,, 
\end{align}
where
\begin{align} \label{eq:generalO}
    \tilde{\CO} & = \CO + U \, d_n + \tilde{W}^{ijk} \, \nabla_i \nabla_j \nabla_k + X^{ij} \, \nabla_i \nabla_j + \tilde{Y}^i \, \nabla_i + Z\,.
\end{align}
Here, $\CO$ is defined in \eqref{eq:Ononp}, which we record here:
\begin{align} \label{eq:CO2}
    \CO & = - d_n^2 - K \, d_n + \frac{1}{N} \, \Box \, N \, \Box \notag \\[2pt]
    & = - d_n^2 - K \, d_n + \Box^2 - 2 \, a^i \, \nabla_{\!i} \, \Box + \bigl( a^i \, a_i - \nabla^i a_i \bigr) \, \Box\,.
\end{align}
Note that $\tilde{W}^{ijk} = \tilde{W}^{ikj}$ and $X^{ij} = X^{ji}$. 
Under the condition
\be \label{eq:matchingr}
    U =
    \tilde{W}^{ijk} = 
    X^{ij} =
    {\tilde{Y}}^i = Z = 0\,,
\ee
the operator in \eqref{eq:generalO} reduces to the one in \eqref{eq:Ononp}. Note that the action in \eqref{eq:genaction} typically breaks the anisotropic Weyl invariance and thus we expect more general operators to appear in the one-loop effective action. One may also introduce a more general operator $V^{ijk\ell} \, \nabla_{\!i} \, \nabla_{\!j} \, \nabla_{\!k} \, \nabla_{\!\ell}$ that replaces $\Box^2$ in the operator $\tilde{\CO}$ defined in \eqref{eq:generalO}. However, since the coefficeint $V^{ijk\ell}$ is dimensionless and should be covariant under the foliation preserving diffeomorphisms, the operator $V^{ijk\ell} \, \nabla_{\!i} \, \nabla_{\!j} \, \nabla_{\!k} \, \nabla_{\!\ell}$ can be reduced to the ones already included in $\tilde{O}$\,. In more complicated theories, a richer structure may arise and require introducing a general $V^{ijk\ell}$. 

We further note that
\be
    \tilde{W}^{ijk} \, \nabla_{\!i} \nabla_{\!j} \nabla_{\!k} \phi = {\tilde{W}}^{(ijk)} \, \nabla_{\!i} \nabla_{\!j} \nabla_{\!k} \phi - \frac{1}{3} \bigl( \tilde{W}^{ij}{}_j - {\tilde{W}}_j{}^{ji} \bigr) \, R\, \nabla_i \phi\,.
\ee
It is therefore convenient to define 
\be \label{eq:WY}
    W^{ijk} \equiv \tilde{W}^{(ijk)}\,,
        \qquad
    Y^i \equiv \tilde{Y}^i - \frac{1}{3} \bigl( \tilde{W}^{ij}{}_j - \tilde{W}_j{}^{ji} \bigr) \, R\,.
\ee
Using \eqref{eq:WY}, we rewrite $\tilde{\CO}$ in \eqref{eq:generalO} as
\be \label{eq:nonminop}
    \tilde{\CO} = \CO + U \, d_n + W^{ijk} \, \nabla_i \nabla_j \nabla_k + X^{ij} \, \nabla_i \nabla_j + Y^i \, \nabla_i + Z\,.
\ee
This change of basis allows us to have a much more succinct result for the heat kernel coefficients.

\begin{figure}[t]
\centering
\includegraphics[scale=0.46]{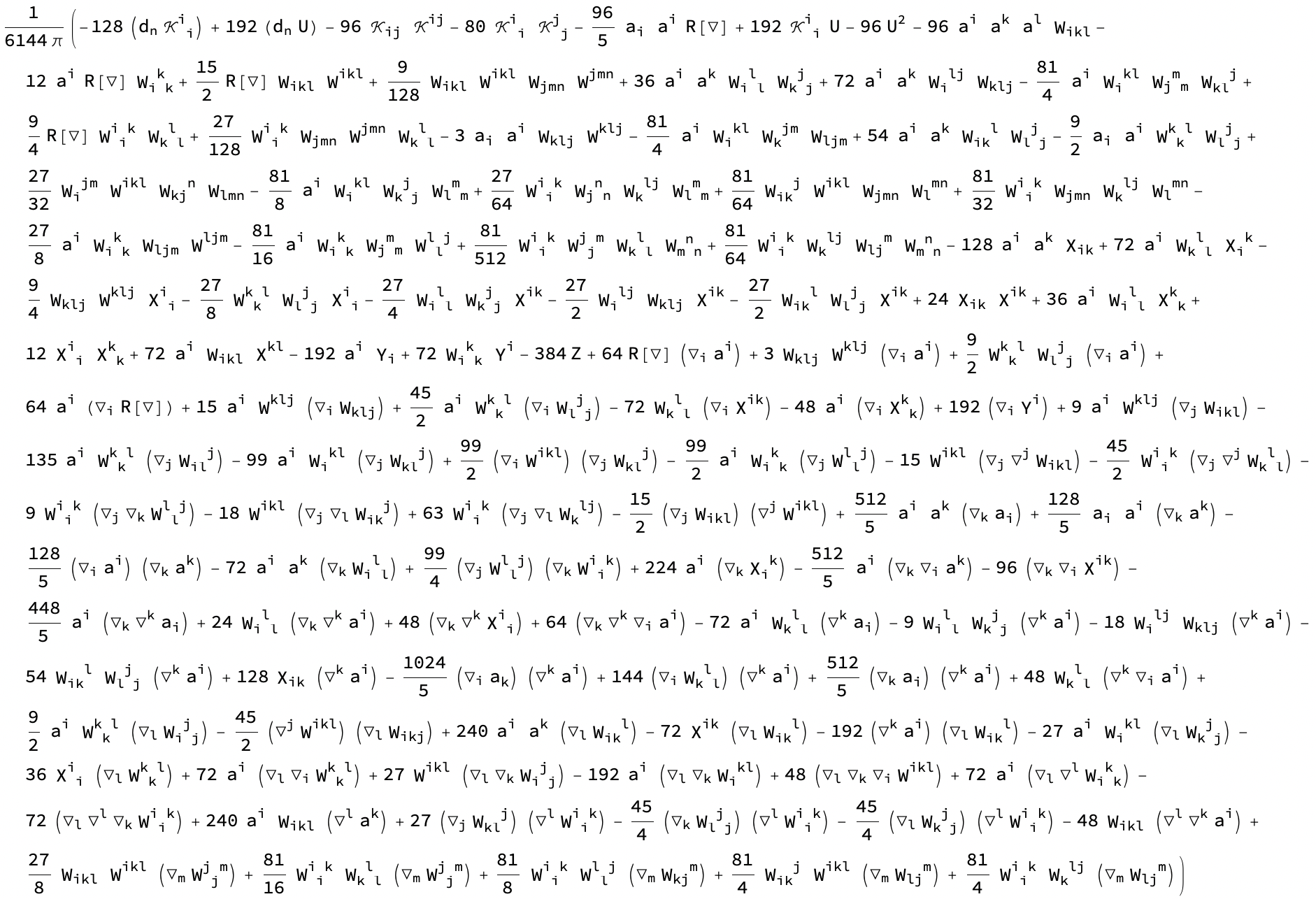}
\caption{Expression of $\tilde{E}^{(4)}$ copied from Mathematica.}
\label{fig:result}
\end{figure}

Following the same procedure described in \S\ref{sec:awals}, we find $\tilde{\SG}^{(0)} = \SG^{(0)}$ and
\begin{equation} \label{eq:SG2}
    \tilde{\SG}^{(2)} = \SG^{(2)} + \frac{1}{512 \pi^{3/2}} \bigl( 16 \, X_{i}{}^{i} - 3 \, W_{ij}{}^{j} \, W^{ik}{}_{k} - 2 \, W_{ijk} \, W^{ijk} \bigr) + \frac{1}{N \sqrt{h}} \, \partial_i f^i,
\end{equation}
where $E^{(2)}$ is given in \eqref{eq:a2exp} and 
\begin{equation}
    f^i = - \frac{3}{64 \pi^{3/2}} \, N \, W^{ij}{}_j\,.   
\end{equation}
A detailed derivation of $\tilde{E}^{(2)}$ can be found in Appendix \ref{app:E2}.
We also find, using xAct \cite{xAct}, the result for ${\tilde{\SG}}^{(4)}$ in Figure \ref{fig:result}. We will transcribe this result in Appendix \ref{app:a4W}. Under the condition \eqref{eq:matchingr}, the result in Figure \ref{fig:result} reduces to $\tilde{\SG}^{(4)} = \SG^{(4)}$.
The one-loop effective action is related to $\tilde{\SG}^{(4)}$ similarly as in \eqref{eq:Gamma1W}, with
\begin{align} \label{eq:Gamma1}
    \Gamma_1 = \frac{1}{2} \lr \log \frac{M^2}{\mu^2} - \gamma^{}_E \rr \int dt \, d^2 \mathbf{x} \, N \sqrt{h} \, {\tilde{\SG}}^{(4)} + \text{finite}.
\end{align}

One particularly useful case is when 
\be
    W^{ijk} = \frac{w}{3} \, \bigl( a^i \, h^{jk} + a^j \, h^{ki} + a^k \, h^{ij} \bigr)\,,
\ee
where $w$ is a constant number. Let us write
\be \label{eq:a4Wo}
    \tilde{\SG}^{(4)} = \SG^{(4)} + \sum_{p = 0}^4 b_p\,,
\ee
where $\SG^{(4)}$ is given in \eqref{eq:a4re} and $b_p$ contains terms of $p$-th order in $W$, with $p = 0\,, \cdots\,, 4$\,.
In this case, the results for $b_1\,, \cdots\,, b_4$ simplify significantly. The expressions for $b_p$\,, $p = 0\,, \cdots\,, 4$ are given below,
\begin{align} \label{eq:simb}
    b_0 & = \frac{1}{256 \pi} \bigl( X_{ij} \, X^{ij} + \tfrac{1}{2} \, X^i{}_i \, X^j{}_j \bigr) - \frac{1}{64 \pi} \bigl( U^2 + 4 Z \bigr ) + \frac{1}{N\sqrt{h}} \bigl( \p_t f^t_0 + \p_i f^i_0 \bigr) \,, \notag \\[2pt]
    b_1 & = - \frac{w}{256 \pi} \, \Bigl[ 4 \, \nabla^j X_{ij} + 3 \, X^j{}_j \, \bigl( \nabla_i - a_i \bigr) + 2 \, X_{ij} \bigl( \nabla^j - 3 \, a^j \bigr) - 4 \, Y_i \Bigr] \, a^i  
    + \frac{w}{N \sqrt{h}} \, \p_i f^i_1\,, \notag \\[2pt]
    b_2 & = \frac{w^2 \, a^i}{3072 \pi} \, \Bigl[ \bigl( 19 \, R - 9 \, X_j{}^j + 48 \, a^2 - 24 \, \nabla_{\!j} a^j \bigr) \, a_i -18 \, \bigl( X_{ij} \, a^j + \Box \, a_i \bigr) \Bigr]
    + \frac{w^2}{N \sqrt{h}} \, \p_i f^i_2\,, \notag \\[2pt]
    b_3 & = \frac{3 \, w^3 \, a^2}{512 \pi} \, \bigl( \nabla^i a_i - 2 \, a^2 \bigr) + \frac{w^3}{N \sqrt{h}} \, \p_i f^i_3\,, \notag \\[2pt]
    b_4 & = \frac{15 \, w^4 \, a^4}{8192\pi}\,,
\end{align}
where
\begin{subequations}
\begin{align}
    f^t_0 & = \frac{\sqrt{h}}{32 \pi} \,  U\,, \\[2pt]
    f^i_0 & = \frac{\sqrt{h}}{32\pi} \Bigl\{ - N^i \, U + N  \ls Y^i + \tfrac{1}{12} \bigl( 8 \,a_j \, X^{ij} - 6 \, \nabla_{\!j} X^{ij} + 3 \, \nabla^i X^j{}_j \bigr) \rs \Bigr\}\,, \\[2pt]
    f^i_1 & = \frac{\sqrt{h}}{384\pi} \, N \Bigl[ \bigl( R + 6 \, a^2 - 4 \, \nabla_{\!j} \, a^j \bigr) \, a^i - 3 \, \nabla^{i} \nabla_{\!j} a^j + 4 \, a^j \, \nabla^{i} a_j \Bigr]\,, \\[2pt]
    f^i_2 & = - \frac{\sqrt{h}}{512\pi} \,N  \Bigl[ \bigl( 2 \, a^j \, \nabla_{\!j} - 7 \, \nabla_{\!j} \, a^j + 9 \, a^2 \bigr) \, a^i \Bigr]\,, \\[2pt]
    f^i_3 & = \frac{3 \sqrt{h}}{1024\pi} \, N \, a^2 a^i\,.
\end{align}
\end{subequations}

Let us now return to the discussion of the coefficient of $a_i \, a^i$ reported in \cite{DOdorico:2015pil}. As we noted at the end of the previous subsection, the operator that they are studying is not the classically Weyl-invariant one. In the language of this subsection, we can write the operator considered in \cite{DOdorico:2015pil} as the Weyl-invariant one plus terms parametrized by $U$, $W$, $X$, $Y$ and $Z$ as
\begin{subequations}
\begin{align}
    U &= W^{ijk} = Z = 0\,, \\
    X^{ij} &= ( \nabla_{\!i} \, a^i - a_i \, a^i ) \, h^{ij} + a^i a^j - 2 \, \nabla^i a^{j}\,, \\
    Y^i &= a_j \nabla^j a^{i} - \Box \, a^i - \frac{1}{2} R \, a^i\,.
\end{align}
\end{subequations}
It is easy to see that the trace of $X^{ij}$ is $X_{i}{}^{i} = - a_i a^i$ and, therefore, the second heat kernel coefficient in \eqref{eq:SG2} simplifies to
\begin{equation}
    \tilde{\SG}^{(2)} = \SG^{(2)} - \frac{1}{32 \pi^{3/2}} \, a_i \, a^i.
\end{equation}
The coefficients reported in Table 1 of \cite{DOdorico:2015pil} factors out a factor of $(4 \pi )^{-(D+1)/2}$. Thus, we would expect the coefficient of $a_i \, a^i$, which they denote by $c_1$, to be
\begin{equation}
    c_1 = - \frac{1}{32 \pi^{3/2}} ( 4 \pi )^{3/2} = - \frac{1}{4}.
\end{equation}
Instead, \cite{DOdorico:2015pil} gets $- \frac{13}{12}$. Given the concerns voiced in \cite{Barvinsky:2017mal} regarding the method used in \cite{DOdorico:2015pil}, a resolution of this discrepancy would be most welcome. 

%%%%%%%%%%%%%%%%%%%%%%%%%%%%%%
%%%%%%%%%%%%%%%%%%%%%%%%%%%%%%

\section{Conclusions} \label{sec:concl}

In this paper we develop a new heat kernel method for calculating the one-loop effective action in Lifshitz theories on a curved background geometry with anisotropic scaling. We tested this method by applying it to the computation of the anisotropic Weyl anomaly for a $(2+1)$-dimensional scalar field theory around a $z=2$ Lifshitz point, and corroborated the results previously found by other heat kernel methods and also by holographic renormalizaiton. In addition, since the form of the anisotropic Weyl anomaly is highly constrained by the Wess-Zumino consistency condition, it serves as a strong check that our method applies to field theories on nonrelativistic background geometries without the projectability condition. We then took one step further and computed the effective action for the most general scalar operators around a $z=2$ fixed point in $2+1$ dimensions. 

One interesting next step is to evaluate the effective action for the most general $(3+1)$-dimensional scalar operator around a $z=3$ fixed point. In particular, the complete cohomologies for anisotropic Weyl anomalies in spacetime dimensions $d \leq 4$ and for dynamical critical exponents $z \leq 3$ are classified in \cite{Arav:2014goa}, and some of the anisotropic Weyl anomalies have essentially been computed in \cite{DOdorico:2015pil, Barvinsky:2017mal}. This can provide us with further checks of our new method, including the terms dependent on $a_i$ (i.e., terms dependent on the spatial variation of the lapse function).  

Another interesting topic is the study of the Weyl anomalies of Ho\v{r}ava gravity with a $U(1)$ extension \cite{Horava:2010zj}. This type of theory was formulated covariantly using torsional Newton-Cartan geometry in \cite{Hartong:2015zia}. In \cite{Christensen:2013lma, Christensen:2013rfa}, boundary Weyl anomalies that take the form of Ho\v{r}ava gravity for bulk torsional Newton-Cartan gravity are derived using the Fefferman-Graham expansion. It will be intriguing to reproduce and extend these results using our new method. 

We emphasize that our method directly applies to operators with spin structures, simply by turning on the bundle structure on the Lifshitz scalar field. This promises many applications to Yang-Mills theory and Ho\v{r}ava gravity. For example, it will be fascinating to generalize the RG flow results in \cite{Barvinsky:2017kob} to non-projectable Ho\v{r}ava gravity in $2+1$ dimensions, which are essential ingredients for understanding the quantum membrane theory at quantum criticality proposed in \cite{Horava:2008ih}.

\acknowledgments

We would like to thank Andrei Barvinsky, Diego Blas, Shira Chapman, Mario Herrero-Valea, and Sergey Sibiryakov for useful discussions. KTG and CMT would like to thank the hospitality of Perimeter Institute for Theoretical Physics, where part of this work was conducted. KTG and ZY would like to thank the hospitality of CERN for stimulating discussions that partly motivated this work. ZY is grateful for the hospitality of Julius-Maximilians-Universit\"{a}t W\"{u}rzburg, where this work was initiated. KTG acknowledges financial support from the Deutsche Forschungsgemeinschaft (DFG, German Research Foundation) under Germany's Excellence Strategy through the W{\"u}rzburg-Dresden Cluster of Excellence on Complexity and Topology in Quantum Matter -- ct.qmat (EXC 2147, project--id 390858490) as well as the Hallwachs-R{\"o}ntgen Postdoc Program of ct.qmat. The work of CMT was supported through a research fellowship from the
Alexander von Humboldt foundation. This research is supported in part by Perimeter Institute for Theoretical Physics. Research at Perimeter Institute is supported in part by the Government of Canada through the Department of Innovation, Science and Economic Development Canada and by the Province of Ontario through the Ministry of Colleges and Universities.

\appendix

%%%%%%%%%%%%%%%%%%%%%%%%%%%%%%%%%%%%
%%%%%%%%%%%%%%%%%%%%%%%%%%%%%%%%%%%%

\section{Procedural Example: Second Heat Kernel Coefficient} \label{app:E2}

Let us show how our method proceeds in complete detail for the calculation of $\SG^{(2)}$. We first introduce some notation. We will use the standard semicolon notation for covariant derivatives. However, keep in mind that because we are using the standard notation, we will also adopt the standard way of ordering the indices: the indices are ordered left-to-right in the order in which the derivatives act on the field, which is right-to-left. This is in contrast to the notation that we introduced for derivatives acting on $\psi$: for example, $\psi_{ijk} = \nabla_{\!i} \nabla_{\!j} \nabla_{\!k} \psi$, which can also be written as $\psi_{;kji}$. We also extend the notation we introduced for spatial derivatives to time derivatives: for example, $\psi_{ni} \equiv d_n \nabla_{\!i} \psi$. So as to reduce the clutter, let us introduce some notation. Firstly, note that the recursion relation for $\sigma^{(2)}$ involves only the operators $\scD^{(0)}$, $\scD^{(1)}$ and $\scD^{(2)}$, but not $\scD^{(3)}$ or $\scD^{(4)}$\,, where $\scD^{(I)}$, $I = 0\,, \cdots\,, 4$ are defined in \eqref{eq:Asigma2}. Let us denote $\scD^{(0)}$ by
\begin{equation}
    \scA \equiv \scD^{(0)}\,.
\end{equation}
This will never contain any dangling derivatives that act on the scalar field. On the other hand, $\scD^{(1)}$ contains up to one derivative acting on the scalar field. So, we write it as
\begin{equation}
    \scD^{(1)} = \scB + \scB^i \nabla_{\!i}\,.
\end{equation}
We will try to be as general as possible until the end, so we will not yet plug in specific expressions for $\scB$ and $\scB^i$, but one can easily do so for the Weyl-invariant case or even the general case.

Next, $\scD^{(2)}$ contains up to two spatial derivatives or one time derivative acting on the scalar field. So, we write it as
\begin{equation}
    \scD^{(2)} = \scC + \scC^i \nabla_i + \scC^{ij} \nabla_{\!i} \nabla_{\!j} + \scC_n d_n\,,
\end{equation}
Again, we will not yet plug in specific expressions for the coefficient functions $\scC$, $\scC^i$, $\scC^{ij}$, and $\scC_n$, but one could certainly do so for whichever operator is of interest.

The $\sigma^{(2)}$ recursion relation now reads
\begin{equation}
    0 = \scA \sigma^{(2)} + \scB^i \sigma_{;i}^{(1)} + \scB \sigma^{(1)} + \scC^{ij} \sigma_{;ji}^{(0)} + \scC_n \sigma_{;n}^{(0)} + \scC^i \sigma_{;i}^{(0)} + \scC \sigma^{(0)}\,.
\end{equation}
From this equation, subtract the contraction of $\scA^{-1} \scB^i$ with the derivative of the $\sigma^{(1)}$ recursion relation:
\begin{equation}
    0 = \scA \sigma_{;i}^{(1)} + \scA_{;i} \sigma^{(1)} + \scB^j \sigma_{;ji}^{(0)} + \scB \, \sigma_{;i}^{(0)} + \scB_{;i}^{j} \, \sigma_{;j}^{(0)} + \scB_{;i} \, \sigma^{(0)} \,.
\end{equation}
This will get rid of the $\sigma_{;i}^{(1)}$ term, leaving us with
\begin{align} \label{eq:Asigma2}
    0 &= \scA \sigma^{(2)} + \bigl( \scB - \scA^{-1} \scA_{;i} \, \scB^i \bigr) \sigma^{(1)} + \bigl( \scC^{ij} - \scA^{-1} \scB^i \scB^j \bigr) \sigma_{;ji}^{(0)} + \scC_n \sigma_{;n}^{(0)} \notag \\
    &\quad + \bigl[ \scC^i - \scA^{-1} \bigl( \scB \scB^i + \scB^j \scB_{;j}^{i} \bigr) \bigr] \sigma_{;i}^{(0)} + \bigl( \scC - \scA^{-1} \scB^i \scB_{;i} \bigr) \sigma^{(0)} \,.
\end{align}
We just keep doing this repeatedly, each time getting rid of the term with the largest number of derivatives acting on $\sigma^{(0)}$, keeping mindful of the fact that $\sigma^{(m)}$ itself contains up to $m$ derivatives acting on $\sigma^{(0)}$ by the recursion relations. For example, next, we would subtract the contraction of $\scA^{-1} ( \scC^{ij} - \scA^{-1} \scB^i \scB^j )$ with the derivatives $\nabla_{\!i} \nabla_{\!j}$ acting on the $\sigma^{(0)}$ defining equation:
\begin{equation} \label{eq:s0ij}
    I_{;ji} = \scA \sigma_{;ji}^{(0)} + 2 \scA_{;(j} \, \sigma_{;i)}^{(0)} + \scA_{;ji} \, \sigma^{(0)}\,.
\end{equation}
We can simultaneously subtract $\scA^{-1} \scC_n$ multiplied by the time derivative of the $\sigma^{(0)}$ defining equation:
\begin{equation}
    I_{;n} = \scA \sigma_{;n}^{(0)} + \scA_{;n} \sigma^{(0)}\,.
\end{equation}
This will remove the terms $\sigma_{;ji}^{(0)}$ and $\sigma_{;n}^{(0)}$ in \eqref{eq:Asigma2} and leaves us with
\begin{align}
    0 &= \scA \sigma_2 + \scA^{-1} ( \scC^{ij} - \scA^{-1} \scB^i \scB^j ) I_{;ji} + \scA^{-1} \scC_n I_{;n} + \bigl( \scB - \scA^{-1} \scA_{;i} \scB^i \bigr) \sigma^{(1)} \notag \\
    &\quad + \bigl[ \scC^i - \scA^{-1} \bigl( \scB \scB^i + \scB^j \scB_{;j}^{i} + 2 \scA_{;j} \scC^{ij} \bigr) + 2 \scA^{-2} \scA_{;j} \scB^j \scB^i \bigr] \sigma_{;i}^{(0)} \notag \\
    &\quad + \bigl[ \scC - \scA^{-1} \bigl( \scB^i \scB_{;i} + \scA_{;ji} \scC^{ij} + \scA_{;n} \scC_n \bigr) + \scA^{-2} \scA_{;ji} \scB^i \scB^j \bigr] \sigma^{(0)}\,.
\end{align}
Next, subtract $\scA^{-1} \bigl( \scB - \scA^{-1} \scA_{;i} \scB^i \bigr)$ multiplied by the $\sigma^{(1)}$ recursion relation
\begin{equation}
    0 = \scA \sigma^{(1)} + \scB^i \sigma_{;i}^{(0)} + \scB \sigma^{(0)} \,,
\end{equation}
which leaves us with
\begin{align}
    0 &= \scA \sigma^{(2)} + \scA^{-1} ( \scC^{ij} - \scA^{-1} \scB^i \scB^j ) I_{;ji} + \scA^{-1} \scC_n I_{;n} \notag \\
    &\quad + \bigl[ \scC^i - \scA^{-1} \bigl( 2 \scB \scB^i + \scB^j \scB_{;j}^{i} + 2 \scA_{;j} \scC^{ij} \bigr) + 3 \scA^{-2} \scA_{;j} \scB^j \scB^i \bigr] \sigma_{;i}^{(0)} \notag \\
    &\quad + \bigl[ \scC - \scA^{-1} \bigl( \scB^2 + \scB^i \scB_{;i} + \scA_{;ji} \scC^{ij} + \scA_{;n} \scC_n \bigr) \notag \\
    &\hspace{1cm} + \scA^{-2} \bigl( \scA_{;i} \scB^i \scB + \scA_{;ji} \scB^i \scB^j \bigr) \bigr] \sigma^{(0)}\,.
\end{align}
Now, we subtract the appropriate term involving $\nabla_i$ acting on the $\sigma^{(0)}$ defining equation to finally get an expression for $\sigma^{(2)}$ that depends only on $\sigma^{(0)}$. Finally, we can replace $\sigma^{(0)}$ with $\scA^{-1} I$:
\begin{align}
    0 &= \scA \sigma_2 + \scA^{-1} ( \scC^{ij} - \scA^{-1} \scB^i \scB^j ) I_{;ji} + \scA^{-1} \scC_n I_{;n} \notag \\
    &\quad + \scA^{-1} \bigl[ \scC^i - \scA^{-1} \bigl( 2 \scB \scB^i + \scB^j \scB_{;j}^{i} + 2 \scA_{;j} \scC^{ij} \bigr) + 2 \scA^{-2} \scA_{;j} \scB^j \scB^i \bigr] I_{;i} \notag \\
    &\quad + \scA^{-1} \bigl[ \scC - \scA^{-1} \bigl( \scB^2 + \scB^i \scB_{;i} + \scA_{;i} \scC^i + \scA_{;ji} \scC^{ij} + \scA_{;n} \scC_n \bigr) \notag \\
    &\hspace{2cm} + \scA^{-2} \bigl( 3 \scA_{;i} \scB^i \scB + \scA_{;ji} \scB^i \scB^j + \scA_{;i} \scB^j \scB_{;j}^{i} + 2 \scA_{;i} \scA_{;j} \scC^{ij} \bigr) \notag \\
    &\hspace{2cm} - 3 \scA^{-3} ( \scA_{;i} \scB^i )^2 \bigr] I\,.
\end{align}
Once we take the coincidence limit, all the terms with derivatives acting on $I$ vanish, $I$ just turns into unity, and $\scA^{-1}$ turns into $G$, the propagator of the theory.\footnote{The coincidence limits of $I_{;n}$ and $I_{;i}$ vanish by definition. The term $I_{;ji}$ is contracted with either $\scC^{ij}$ or $\scB^i \scB^j$, both of which are symmetric. Thus, we can symmetrize the derivatives and write $I_{;(ji)}$ the coincidence limit of which also vanishes by definition.} Thus, the final expression for the coincidence limit of $\sigma^{(2)}$ is
\begin{align} \label{eq:s2cl}
    \sigma^{(2)} \bigr |_{x=x_0} &= - G^2 \bigl[ \scC - G \bigl( \scB^2 + \scB^i \scB_{;i} + \scA_{;i} \scC^i + \scA_{;ji} \scC^{ij} + \scA_{;n} \scC_n \bigr) \notag \\
    &\hspace{1cm} + G^2 \bigl( 3 \scA_{;i} \scB^i \scB + \scA_{;ji} \scB^i \scB^j + \scA_{;i} \scB^j \scB_{;j}^{i} + 2 \scA_{;i} \scA_{;j} \scC^{ij} \bigr) \notag \\
    &\hspace{1cm} - 2 G^3 ( \scA_{;i} \scB^i )^2 \bigr] \bigr|_{x=x_0} \,.
\end{align}
We emphasize that this is the general expression for the coincidence limit of $\sigma^{(2)}$ for \emph{any} $z=2$ differential operator acting on a scalar field. For a specific operator, we can compute the coincidence limits of the operator coefficients remaining above.

Explicit expressions for the operator coefficients for the general $z=2$ operator written in the specific form in \eqref{eq:nonminop}, with the Weyl-invariant operator defined in \eqref{eq:Ononp}, are
\begin{subequations}
\begin{align}
    \scA &= ( \chi_n )^2 + ( \psi_i \psi^i )^2 - \lambda \,, \\
    \scB &= 2 \chi_n \psi_n - 2i \psi^i \psi_i \psi_{j}{}^{j} - 4i \psi^i \psi^j \psi_{ij} + 2ia^i \psi_i \psi_j \psi^j -i W^{ijk} \psi_i \psi_j \psi_k \,, \\
    \scB^i &= -4i \psi_j \psi^j \psi^i\,, \\
    \scC &= - i ( \chi_{nn} + K \chi_n ) + ( \psi_n )^2 - ( \psi_{i}{}^{i} )^2 - 2 \psi_{ij} \psi^{ij} - 2 \psi^i \bigl( \psi^{j}{}_{ji} + \psi_{ij}{}^{j} \bigr) \notag \\
    &\hspace{1cm} + ( \nabla \cdot a - a^2 ) \psi_i \psi^i + 2a^i ( 2 \psi_{ij} \psi^j + \psi_i \psi_{j}{}^{j} ) \notag \\
    &\hspace{1cm} + i U \chi_n - 3W^{ijk} \psi_i \psi_{jk} - X^{ij} \psi_i \psi_j\,, \\
    \scC^i &= -4 ( 2 \psi^{ij} \psi_j + \psi^i \psi_{j}{}^{j} ) + 2 ( 2 a_j \psi^j \psi^i + a^i \psi_j \psi^j ) - 3W^{ijk} \psi_j \psi_k\,, \\
    \scC^{ij} &= -2 ( 2 \psi^i \psi^j + \psi_k \psi^k h^{ij} )\,, \\
    \scC_n &= -2i \chi_n\,.
\end{align}
\end{subequations}
Taking one derivative of $\scA$ gives
\begin{equation}
    \scA_{;i} = 2 \, \chi_n \chi_{in} + 4 \, \psi_k \psi^k \psi^j \psi_{ij}\,.
\end{equation}
Since $\psi$ is a scalar, $\nabla_{\!i} \nabla_{\!j} \, \psi = \nabla_{\!j} \nabla_{\!i} \, \psi$, and thus the defining relation $\nabla_{\!(i} \nabla_{\!j)} \psi \bigr|_{x=x_0} = 0$ implies 
\begin{equation}
    \psi_{ij} \bigr|_{x=x_0} = 0\,.
\end{equation}
Using the commutation relation $[d_n , \nabla_{\!i} ] \chi = -a_i \chi_n$ and the defining relation $\chi_n \bigr|_{x=x_0} = \nu$ and $\chi_{ni} \bigr|_{x=x_0} = 0$, we find
\begin{equation}
    \chi_{in} \bigr|_{x=x_0} = \nu a_i \,.
\end{equation}
Therefore,
\begin{equation}
    \scA_{;i} \bigr|_{x=x_0} = 2 \nu^2 a_i\,.
\end{equation}
Taking another derivative we can immediately ignore any terms with two spatial derivatives acting on $\psi$, which we denote by $\ldots$:
\begin{equation}
    \scA_{;ji} = 2 \chi_n \chi_{ijn} + 2 \chi_{in} \chi_{jn} + 4 \psi_{\ell} \psi^{\ell} \psi^k \psi_{ijk} + \ldots \,.
\end{equation}
Using the commutation relations,
\begin{align}
    \chi_{ijn} &= \chi_{inj} - \nabla_{\!i} [ d_n , \nabla_{\!j} ] \chi \notag \\
    &= \chi_{nij} - [ d_n , \nabla_{\!i} ] \chi_j + ( a_j \, \chi_n )_{;i} \notag \\
    &= \chi_{nij} + a_i \, \chi_{nj} + M^{k}{}_{ij} \, \chi_k + a_j \, \chi_{in} + a_{j;i} \, \chi_n \,.
\end{align}
The first three terms vanish in the coincidence limit and the rest give
\begin{equation}
    \chi_{ijn} \bigr|_{x=x_0} = \nu ( a_i a_j + a_{j;i} )\,.
\end{equation}
Thus,
\begin{equation}
    \scA_{;ji} \bigr|_{x=x_0} = 2 \nu^2 ( 2a_i a_j + a_{j;i} )\,.
\end{equation}
Next, let us take coincidence limit of $\scB$ and $\scB^i$:
\begin{subequations}
\begin{align}
    & \scB \bigr|_{x=x_0} = iq^i q^j q^k ( 2a_i h_{jk} - W_{ijk} ) \,, \\
    & \scB^i \bigr|_{x=x_0} = -4i | \bq |^2 q^i \,.
\end{align}
\end{subequations}
Next, we take one derivative of $\scB$. We will denote any terms containing one time or two space derivatives on $\psi$ by $\ldots$ since we know that these vanish in the coincidence limit:
\begin{equation}
    \scB_{;i} = 2 \chi_n \psi_{in} - 2i \psi_k \psi^k \psi_{ij}{}^{j} - 4i \psi^j \psi^k \psi_{ijk} + i ( 2a_{j;i} h_{k \ell} - W_{jk \ell ;i} ) \psi^j \psi^k \psi^{\ell} + \ldots \,.
\end{equation}
Again, using the commutation relations, we have
\begin{equation}
    \psi_{in} \bigr|_{x=x_0} = \bigl( \psi_{ni} + a_i \psi_n \bigr) \bigr|_{x=x_0} = 0\,.
\end{equation}
Using the commutator of two covariant spatial derivatives, we can write
\begin{equation}
    \frac{1}{2} \psi_{(ijk)} = \psi_{ijk} + \psi_{jki} + \psi_{kij} = 3 \psi_{ijk} + 2 R^{\ell}{}_{(k|i|j)} \psi_{\ell} \,.
\end{equation}
In the coincidence limit, the left hand side vanishes by definition, and we find
\begin{equation}
    \psi_{ijk} \bigr|_{x=x_0} = - \frac{2}{3} q^{\ell} R_{\ell (k |i| j)} \,.
\end{equation}
In 2 dimensions, this simplifies to 
\begin{equation}
    \psi_{ijk} \bigr|_{x=x_0} = \frac{1}{3} R ( h_{i (j} q_{k)} - h_{jk} q_i )\,.
\end{equation}
Plugging these back in $\scB_{;i}$ gives
\begin{equation}
    \scB_{;i} \bigr|_{x=x_0} = \frac{2i}{3} R | \bq |^2 q_i + i q^j q^k q^{\ell} \bigl( 2a_{j;i} h_{k \ell} - W_{jk \ell ;i} \bigr)\,.
\end{equation}
Since $\scB^i$ contains only single derivatives on $\psi$, the coincidence limit of $\scB_{;i}^{j}$ simply vanishes:
\begin{equation}
    \scB_{;i}^{j} \bigr|_{x=x_0} = 0\,.
\end{equation}
Next, we must take coincidence limits of the $\scC$ operators:
\begin{subequations}
\begin{align}
    & \scC \bigr|_{x=x_0} = i \nu (U-K) + \biggl( \nabla \cdot a - a^2 - \frac{2}{3} R \biggr) | \bq |^2 - k_i k_j X^{ij} \,, \\
    & \scC^i \bigr|_{x=x_0} = 2 \bigl( 2 a_j q^j q^i + a^i | \bq |^2 \bigr) - 3W^{ijk} q_j q_k \,, \\
    & \scC^{ij} \bigr|_{x=x_0} = -2 \bigl( 2 q^i q^j + | \bq |^2 h^{ij} \bigr) \,, \\
    & \scC_n \bigr|_{x=x_0} = -2i \nu \,.
\end{align}
\end{subequations}
All that is left is to plug in these coincidence limits into the general expression for $\sigma^{(2)}$ in \eqref{eq:s2cl}. For example, setting $U = W^{ijk} = X^{ij} = 0$ gives the expression for $\sigma^{(2)}$ in the case of the Weyl-invariant operator, which we write out in \eqref{eq:s2weyl}. This is then integrated over frequency and momentum to get the heat kernel coefficien $\SG^{(2)}$. The result in the Weyl-invariant case is in \eqref{eq:a2exp} and the result in the general case is in \eqref{eq:SG2}.

The process is exactly the same for the higher-order heat kernel coefficients. The computation will clearly get much more tedious, but is simple in principle and amenable to automation.

%%%%%%%

\section{Expression of \texorpdfstring{${\tilde{\SG}}^{(4)}$}{a4} for General Scalar Operators} \label{app:a4W}

In this appendix, we present the general results for ${\tilde{\SG}}^{(4)}$ in \eqref{eq:Gamma1}, transcribing from Figure \ref{fig:result}. We write
\be 
    \tilde{\SG}^{(4)} = \SG^{(4)} + \sum_{p = 0}^4 b_p\,,
\ee
where $\SG^{(4)}$ is given in \eqref{eq:a4re} and $b_p$\,, $p = 0\,, \cdots\,, 4$ contain terms of $p$-th order in $W$\,. We collect the results for $b_0\,, \cdots\,, b_4$ below:

\vspace{3mm}

\noindent \textbf{Terms Constant in $W$: Coefficient $b_0$\,.} This result already appears in \eqref{eq:simb}, which we write below again for completeness:
\begin{align} \label{eq:b0}
    b_0 & = \frac{1}{256 \pi} \bigl( X_{ij} \, X^{ij} + \tfrac{1}{2} \, X^i{}_i \, X^j{}_j \bigr) - \frac{1}{64 \pi} \bigl( U^2 + 4 Z \bigr ) + \frac{1}{N\sqrt{h}} \bigl( \p_t f^t + \p_i f^i \bigr) \,,
\end{align}
where
\begin{subequations}
\begin{align}
    f^t & = \frac{1}{32 \pi} \, \sqrt{h} \,  U\,, \\[2pt]
    f^i & = \frac{1}{32\pi} \Bigl\{ - \sqrt{h} \, N^i \, U + N  \ls Y^i + \tfrac{1}{12} \bigl( 8 \,a_j \, X^{ij} - 6 \, \nabla_{\!j} X^{ij} + 3 \, \nabla^i X^j{}_j \bigr) \rs \Bigr\}\,. 
\end{align}
\end{subequations}

\vspace{3mm}

\noindent \textbf{Terms Linear in $W$: Coefficient $b_1$\,.} We write $b_1$ as
\begin{align} \label{eq:b1}
    b_1 & = \sum_{q=0}^3 b_{1,q}\,,
\end{align}
where $b_{1,\,q}$ includes all terms that are linear in $W$ and that contain $q$ derivatives acting on $W$. 
We find that
\begin{subequations}
\begin{align}
    b_{1,0} & = 
    - \frac{1}{256\pi} \, \Bigl[ \bigl( 4 \, a_j \, a_k - 10 \, \nabla_{\!j} \, a_k - 3 \, X_{jk} + 2 \, \nabla_{\!j} \nabla_{\!k} \bigr) a_i \Bigr] \, W^{ijk} \notag \\[2pt]
    & \quad + \frac{1}{512 \pi} \, \Bigl[ \bigl( 2 \, \Box - R \bigr) \, a_i - 2 \, \bigl( 3 \, a_{k} \, \nabla_{\!i} - 2 \, \nabla_{\!i} \nabla_{\!k} \bigr) \, a^k \notag \\[2pt]
    & \hspace{2.8cm} + 3 \, a_i \, X^k{}_k + 6 \, \bigl( a_k - \nabla_k \bigr) X_{i}{}^k + 6 \, Y_i \Bigr] W^{ij}{}_j\,, \\[2pt]
    b_{1,1} & = \frac{1}{512\pi} \Bigl[ 6 \, \bigl( 2 \, \nabla^i a^j - a^i \, a^j \bigr) \, \nabla_i W_{jk}{}^k - 3 \, X^i{}_i \, \nabla_{\!j} W^{jk}{}_k \notag \\[2pt]
    & \hspace{2.8cm} + 2 \, \bigl( 10 \, a^i \, a^j - 8 \, \nabla^i a^j - 3 \, X^{ij} \bigr) \, \nabla_{\!k} W_{ij}{}^k \Bigr]\,, \\[2pt]
    b_{1,2} & = \frac{1}{256 \pi} \, a^i \, \bigl( 3 \, \nabla_{\!k} \nabla_{\!i} W_j{}^{jk} - 8 \, \nabla_{\!j} \nabla_{\!k} W_i{}^{jk} + 3 \, \Box W_{ij}{}^j \bigr)\,, \\[2pt]
    b_{1,3} & = \frac{1}{256 \pi} \bigl( 2 \, \nabla_{\!i} \nabla_{\!j} \nabla_{\!k} W^{ijk} - 3 \, \Box \nabla_{\!i} W^{ij}{}_j \bigr)\,.
\end{align}
\end{subequations}

\vspace{3mm}

\noindent \textbf{Terms Quadratic in $W$: Coefficient $b_2$\,.} We start with a classification of possible terms that are quadratic in $W$\,, containing no derivatives acting on any of the $W$'s. The result of $b_2$ will be given later in \eqref{eq:b2}.
We will need the scalars (in which all six indices in the two factors of $W$ combined are contracted into three pairs) and the matrices (the ones with two free, un-conctracted, indices). 
It will also be convenient to define the \emph{multiplicity} of a particular term. The multiplicity of a term is determined by all possible ways of contracting the six indices in the pair of $\zw$'s to produce the term.
The full classification of all scalars and matrices with their multiplicity numbers are listed as follows:
\begin{subequations}
\begin{align}
    & \zs^{(1)} 
    = \zw_{ik}{}^{k} \, \zw^{i \ell}{}_{\ell} &%
    & \text{multiplicity} = 9\,, \\[2pt]
    & \zs^{(2)} 
    = \zw_{ik \ell} \, \zw^{ik \ell} &%
    & \text{multiplicity} = 6\,, \\[2pt]
    & \zm^{(1)}_{ij} 
    = \zw_{ik}{}^{k} \, \zw_j{}^{\ell}{}_{\ell} &%
    & \text{multiplicity} = 9\,, \\[2pt]
    & \zm^{(2)}_{ij}  
    = \zw_{ik \ell} \, \zw_j{}^{k \ell} &%
    & \text{multiplicity} = 18\,, \\[2pt]
    & \zm^{(3)}_{ij}  
    = \zw^{k \ell}{}_{\ell} \, \zw_{ijk} &%
    & \text{multiplicity} = 36\,, 
\end{align}
\end{subequations}
We can then define the sum of all possible scalars (counting multiplicity):
\be \label{eq:zp}
    \zp = 3 \, \bigl(3 \, \zs^{(1)} + 2 \, \zs^{(2)} \bigr)\,. 
\ee
Similarly, the sum of all possible matrices (counting multiplicity) is
\begin{equation} \label{eq:zq}
    \zq = 9 \, \bigl( \zm^{(1)} + 2 \, \zm^{(2)} + 4 \, \zm^{(3)} \bigr)\,.
\end{equation}
Using the notation introduced above, we write 
\be \label{eq:b2}
    b_2 = \sum_{q=0}^2 b_{2,q}\,,
\ee
where $b_{2,\,q}$ includes all terms that are quadratic in $W$ and that contain $q$ derivatives acting on $W$'s. 
We have
\begin{subequations}
\begin{align}
    b_{2,0} & =  \frac{2^{-14}}{3\pi} \Bigl[ \bigl( 10 \, R - 4 \, a^2 + 4 \, \nabla^i a_i -3 \, X^i{}_i \bigr) \, \zp + 2 \, \bigl( 16 \, a^i \, a^j - 4 \, \nabla^i a^j - 3 \, X^{ij} \bigr) \, \zq_{ij} \notag \\[2pt]
    &  \hspace{3.8cm} - 36 \, \bigl( 2 \, R \, h^{ij} + 20 \, a^i \, a^j + 4 \, \nabla^i a^j - 3 \, X^{ij} \bigr) \, \zm^{(3)}_{ij} \Bigr]\,, \\[2pt]
    b_{2,1} & = \frac{2^{-12}}{\pi} \, a^i \, \Bigl[ 2 \, W^{jk\ell} \, \bigl( 5 \, \nabla_{\!i} W_{jk\ell} + 3 \, \nabla_{\!j} W_{ik\ell} \bigr) - 2 \, W_{ijk} \, \bigl( 33 \, \nabla_{\!\ell} W^{jk\ell} + 9 \, \nabla^j W^{k\ell}{}_{\ell} \bigr) \notag \\[2pt]
    & \hspace{1.6cm} - 33 \, W_{ik}{}^k \, \nabla_{\!j} W^{j\ell}{}_\ell + 3 \, W^{jk}{}_k \bigl( 5 \, \nabla_{\!i} W_{j\ell}{}^\ell + \nabla_{\!j} W_{i\ell}{}^\ell - 30 \, \nabla^{\ell} W_{ij\ell} \bigr) \Bigr]\,, \\[2pt]
    b_{2,2} & = - \frac{2^{-12}}{\pi} \, \Bigl[3 \, W^i{}_{ij} \bigl( 5 \, \Box W^{jk}{}_k + 2 \, \nabla_{\!k} \nabla^{j} W^{k\ell}{}_\ell - 14 \, \nabla_{\!k} \nabla_{\!\ell} W^{jk\ell} \bigr) \notag \\[2pt]
    & \hspace{4cm} + 2 \, W^{ijk} \bigl( 5 \, \Box W_{ijk} + 6 \, \nabla_{\!\ell} \nabla_{k} W_{ij}{}^\ell - 9 \, \nabla_{\!j} \nabla_{\!k} W_{i\ell}{}^\ell \bigr) \Bigr] \notag \\[2pt]
    & \quad + \frac{2^{-13}}{\pi} \, \Bigl[ 33 \, \bigl( 2 \, \nabla_{\!i} W^{ik\ell} \, \nabla^{j} W_{jk\ell} + \nabla_{\!i} W^{ik}{}_k \, \nabla_{\!j} W^{j\ell}{}_\ell \bigr) \notag \\[2pt]
    & \hspace{4cm} - 3 \, \nabla^i W^{jk}{}_k \, \bigl( 5 \, \nabla_{\!i} W_{j\ell}{}^\ell + 5 \, \nabla_{\!j} W_{i\ell}{}^\ell - 12 \, \nabla_{\!\ell} W^\ell{}_{ij} \bigr) \notag \\[2pt]
    & \hspace{4cm} -10 \, \nabla^\ell W^{ijk} \, \bigl( 3 \, \nabla_k W_{ij\ell} + \nabla_{\!\ell} W_{ijk} \bigr) \Bigr]\,.
\end{align}
\end{subequations}

\vspace{3mm}

\noindent \textbf{Terms Cubic in $W$: Coefficient $b_3$\,.} We start with a classification of possible terms that are cubic in $W$\,, containing no derivatives acting on any of the $W$'s. The result of $b_3$ will be given later in \eqref{eq:b3}. We only need the vectors:
\begin{subequations}
\begin{align}
    & \zv^{(1)}_i 
    = \zs^{(1)} \, \zw_i{}^{\ell}{}_{\ell} &%
    & \text{multiplicity} = 81\,, \\[2pt]
    & V^{(2)}_i 
    = \zs^{(2)} \, \zw_i{}^{\ell}{}_{\ell} &%
    & \text{multiplicity} = 54\,, \\[2pt]
    & \zv^{(3)}_i 
    = \zm^{(1)}_{k\ell} \, \zw_i{}^{k\ell} &%
    & \text{multiplicity} = 162\,, \\[2pt]
    & \zv^{(4)}_i 
    = \zm^{(2)}_{k\ell} \, \zw_i{}^{k\ell} &%
    & \text{multiplicity} = 324\,, \\[2pt]
    & \zv^{(5)}_{i} 
    = \zm^{(3)}_{k\ell} \, \zw_i{}^{k\ell} &%
    & \text{multiplicity} = 648\,.
\end{align}
\end{subequations}
Here, the multiplicity of a term is determined by all possible ways of contracting the eight indices in the three $W$'s to produce the term. 
Let $\zv$ be the weighted sum of all of the above terms (including multiplicity):
\begin{align} \label{eq:zv}
    \zv = 3 \, \bigl( 27 \, V^{(1)} + 18 \, V^{(2)} + 54 \, V^{(3)} + 108 \, V^{(4)} + 216 \, V^{(5)} \bigr) \,.
\end{align}
Using the notation introduced above, we write 
\be \label{eq:b3}
    b_{3} = b_{3,\,0} + b_{3,\,1}\,,
\ee
where $b_{3,\,q}$\,, $q=0\,, 1$ includes all terms that are cubic in $W$ and that contain $q$ derivatives acting on $W$'s. We have
\begin{subequations}
\begin{align}
    b_{3,0} & = - \frac{2^{-15}}{3 \pi} \, a_i \, \big( V^i - 324 \,  \zm^{(3)}_{jk} \, \zw^{ijk} \bigr)\,, \label{eq:b30} \\[10pt]
    b_{3,1} & = \frac{3}{2^{15} \pi} \, \bigl( P \, h_{ij} + 2 \, Q_{ij} - 36 \, M^{(3)}_{ij} \bigr) \, \nabla_k W^{ijk}.
\end{align}
\end{subequations}

\vspace{3mm}

\noindent \textbf{Terms Quartic in $W$: Coefficient $b_4$\,.} We start with a classification of possible terms that are quartic in $W$\,, containing no derivatives acting on any of the $W$'s:
\begin{subequations}
\begin{align}
    & \zcw^{(1)} = S^{(1)} \, S^{(1)}
        &
    & \text{multiplicity} = 81\,, \\[2pt]
    & \zcw^{(2)} = S^{(1)} \, S^{(2)}
        &
    & \text{multiplicity} = 108\,, \\[2pt]
    & \zcw^{(3)} = S^{(2)} \, S^{(2)}
        &
    & \text{multiplicity} = 36\,, \\[2pt]
    & \zcw^{(4)} = \tr \bigl( M^{(1)} \, M^{(2)} \bigr) 
        &
    & \text{multiplicity} = 648\,, \\[2pt]
    & \zcw^{(5)} = \tr \bigl( M^{(1)} \, M^{(3)} \bigr) 
        &
    & \text{multiplicity} = 216\,, \\[2pt]
    & \zcw^{(6)} = \tr \bigl( M^{(2)} \, M^{(2)} \bigr) 
        &
    & \text{multiplicity} = 648\,, \\[2pt]
    & \zcw^{(7)} = \tr \bigl( M^{(2)} \, M^{(3)} \bigr) 
        &
    & \text{multiplicity} = 1296\,, \\[2pt]
    & \zcw^{(8)} = W_{ijk} \, W^{i\ell}{}_{m} \, W^j{}_{\ell n} \, W^{kmn}
        &
    & \text{multiplicity} = 432\,.
\end{align}
\end{subequations}
Here, the multiplicity of a term is determined by all possible ways of contracting the twelve indices in the four $W$'s to produce the term. 
Let $\zcw$ be the weighted sum of all of the above terms (including multiplicity):
\begin{align} \label{eq:zcw}
    \zcw = 9 \, \bigl( 9 \, \zcw^{(1)} + 12 \, \zcw^{(2)} + 4 \, \zcw^{(3)} & + 72 \, \, \zcw^{(4)} + 24 \, \zcw^{(5)} \notag \\[2pt]
        & + 72 \, \zcw^{(6)} + 144 \, \zcw^{(7)} + 48 \, \zcw^{(8)} \bigr)\,.
\end{align}
In terms of $W$ in \eqref{eq:zcw}, we find 
\be \label{eq:b4}
    b_4 = \frac{2^{-20}}{3 \pi} \, \zcw\,.
\ee
Note that there is an interesting relation between $b_3$ and $b_4$\,,
\begin{align}
    & \quad b_4 \Bigl( W_{ijk} \rightarrow w \, W_{ijk} - \tfrac{2}{3} \, \bigl( a_i \, h_{jk} + a_j \, h_{ki} + a_k \, h_{ij} \bigr) \Bigr) \notag \\[2pt]
    & = w^4 \, b_4 (W) + w^3 \, b_{3, 0} (a\,, W) + \CO (w^2)\,, 
\end{align}
where $b_3$ and $b_4$ are given in \eqref{eq:b30} and \eqref{eq:b4}, respectively.

%%%%%%%%%%%%%%%%%%%%%%%%%%%%%%%%%%%%%%%%
% Appendix C
%%%%%%%%%%%%%%%%%%%%%%%%%%%%%%%%%%%%%%%%

\section{Sign Conventions in the Literature} \label{app:dis}

As pointed out in \cite{DOdorico:2015pil}, the heat kernel coefficient of the monomials built out of the spatial curvature in a Lifshitz theory in which the spatial derivative part of the operator is essentially just the $z$-th power of the isotropic ($z=1$) case can be expressed as some number (a Mellin integral of some sort) times the $z=1$ result. In the case of $z=D=2$, for the $R^3$ monomial, which shows up in $E^{(6)}$, the appropriate Mellin integral evaluates to $-2$, which is the number reported in the $R^3$ entry of the $z=D=2$ column of their Table 1 (they denote the number of spatial dimensions by $d$). The number $\frac{1}{756}$ to the left of that, in the $a_{2n}$ column, is supposed to be the $z=1$ coefficient of $R^3$ in two spacetime dimensions and the $z=2$ coefficient is simply the product of these two numbers: $-2 \times \frac{1}{756} = - \frac{1}{378}$. However, the result we get using our method is $- \frac{1}{315} = -2 \times \frac{1}{630}$. We agree with the Mellin integral result of $-2$, but we disagree with the $z=1$ coefficient of $R^3$: we get $\frac{1}{630}$ instead of $\frac{1}{756}$. 

Due to this discrepancy, we were led to examine the relativistic heat kernel literature. There are numerous different sign conventions being used in the literature, which can easily cause confusion. We will clarify some of these conventions, which we encountered in the course of verifying the heat kernel coefficients in the relativistic case.

Firstly, we can confirm the correctness of the coefficients calculated by Gilkey \cite{Gilkey:1975iq} in arbitrary dimensions for the terms that contain only Riemann curvatures and their covariant derivatives. We reproduced these coefficients using Gusynin's method reviewed in Section \ref{sec:chk}. One just needs to be mindful of the fact that Gilkey's definition of the Riemann tensor is negative of the one we use here in \eqref{eq:riemanntensor}. The coefficients $E_0$, $E_2$, and $E_4$ are reported on page 610, just after Theorem 2.2, and $E_6$ is given in Theorem 4.1 on page 613. Once evaluated in two dimensions with constant curvature, with the Ricci and Riemann tensors related to the Ricci scalar as in \eqref{eq:riemann2d}, and where the conventional sign for curvature is used (instead of the one used by Gilkey), one derives the result $(4 \pi ) E_6 = \frac{1}{630} R^3$. This factor of $\frac{1}{630}$ is precisely the same one as we get from our aforementioned $z=2$ computation.

In \cite{Vassilevich:2003xt}, Vassilevich calls these coefficients $a_0$ to $a_6$ and reports them in Eqns. (4.26-29). Here, one has to be careful keeping track of the sign conventions as well. Vassilevich defines the Riemann tensor in his Eqn. (2.6), which is negative of the one we use here in \eqref{eq:riemanntensor}. However, his definition of the Ricci tensor and Ricci scalar coincide with ours. Keeping these sign conventions in mind, the coefficients in Vassilevich are consistent with those in Gilkey. The same results are found in App. B and C of \cite{Gustavsson:2019efu} by independent calculation by Gustavsson using the same sign conventions as in Vassilevich.\footnote{We are grateful to Andreas Gustavsson for personal communications that led us to realize our earlier misunderstanding of their sign conventions.}

In \cite{Groh:2011dw}, the authors independently computed $E_6$, though they denote the result by $\overline{A}_3$ in their Eqn. (2.16) (the overline indicates the coincidence limit because this work studies the so-called ``off-diagonal'' heat kernel in which the coincidence limit $x \rightarrow x_0$ may or may not be taken). We focus on the pieces that depend only on the spatial curvature and its derivatives (not on the endomorphism $E$ or the field strength $F_{\mu\nu}$ of the vector bundle describing the non-metric field content of the field theory). Note that their definitions of the Riemann tensor, Ricci tensor, and Ricci scalar have the same sign as ours. The comparison with Gilkey is complicated slightly by the fact that the authors chose to use a slightly different basis for the terms involving two derivatives and two factors of curvature: where Gilkey uses the term $R^{\mu\nu} \nabla^{\rho} \nabla_{\!\mu} R_{\rho\nu}$, the authors of \cite{Groh:2011dw} instead use $R^{\mu\nu} \nabla_{\!\mu} \nabla_{\!\nu} R$. However, we can use the commutator of two covariant derivatives acting on a tensor as well as the second Bianchi identity and the symmetries of the Riemann tensor to write
\begin{align} \label{eq:COB}
    R^{\mu\nu} \nabla^{\rho} \nabla_{\!\mu} R_{\rho\nu} 
    &= \frac{1}{2} R^{\mu\nu} \nabla_{\!\mu} \nabla_{\!\nu} R - R_{\mu\nu} R^{\mu\lambda} R_{\nu}{}^{\lambda} + R^{\mu\nu} R^{\rho\lambda} R_{\mu\rho\nu\lambda} \,.
\end{align}
Thus, we can convert Gilkey's result to the basis used in \cite{Groh:2011dw}, remembering to switch the sign of any terms that are linear or cubic in curvature. We must also keep in mind that $\Delta$ is defined in \cite{Groh:2011dw} with a minus sign: $\Delta = - g^{\mu\nu} \nabla_{\mu} \nabla_{\nu}$. Doing so, we find just one discrepancy in the expression $\overline{A}_3$ in Eqn. (2.16) of \cite{Groh:2011dw}: the coefficient of $R ( \Delta R )$ should be $- \frac{1}{180}$, not $- \frac{1}{280}$, which may just be a misprint.

The coefficients of the terms cubic in curvature contain slight differences (in addition to the overall sign difference) between \cite{Gilkey:1975iq} and \cite{Groh:2011dw} because of the change of basis \eqref{eq:COB}, but they \emph{are} consistent. In particular, in two dimensions with constant curvature, one again gets the result $( 4 \pi ) E_6 = \frac{1}{630} R^3$.

More recently, Kluth and Litim \cite{Kluth:2019vkg} computed up to the $R^3$, $R^4$, and $R^5$ coefficients in $E_6$, $E_8$, and $E_{10}$, respectively, in 2 to 6 dimensions for a sphere (actually for any maximally symmetric space). This is done using quite a different method, which leverages simplifications specific to maximal symmetry from the beginning. Sure enough, the coefficient of $\frac{1}{4 \pi} R^3$ in two dimensions is $\frac{1}{630}$.

%%%%%%%%%%%%%%%%%%%%%%%%%%%%%%%%%%%%%%%%
%%%%%%%%%%%%%%%%%%%%%%%%%%%%%%%%%%%%%%%%

\newpage

\bibliographystyle{JHEP}
\bibliography{hklt}

\end{document}